\documentclass[sigconf,balance=false]{acmart}

\usepackage{popets}
\setcopyright{popets}
\copyrightyear{YYYY}
\acmYear{YYYY}
\acmVolume{YYYY}
\acmNumber{X}
\acmDOI{XXXXXXX.XXXXXXX}
\acmISBN{}
\acmConference{Proceedings on Privacy Enhancing Technologies}
\usepackage[utf8]{inputenc}
\usepackage{xspace}
\usepackage{enumitem}
\usepackage{bookmark}
\usepackage{arydshln} 
\usepackage{soul} 
\usepackage[textwidth=3.7cm]{todonotes}
\usepackage{multirow}
\usepackage{subfigure}
\usepackage{adjustbox}
\usepackage{refcount}
\usepackage{footnote}

\usepackage{tikz}
\usetikzlibrary{positioning}
\usetikzlibrary{patterns}
\usetikzlibrary{fit}
\usetikzlibrary{shapes.geometric}
\tikzstyle{block} = [draw, rectangle, minimum height=3em, minimum width=3em, text width=3em, text centered, inner sep=0pt]
\tikzstyle{content} = [draw, rectangle, minimum height=5em, minimum width=5em, text centered, text width=5em, inner sep=0pt]
\tikzstyle{textblock} = [text width=13em]
\tikzstyle{box} = [draw, rectangle]
\tikzstyle{token} = [regular polygon, draw, minimum size=3em, fill=red!30]

\definecolor{darkgreen}{rgb}{0,0.6,0}

\newcommand{\ignore}[1]{}

\usepackage{amsmath,amsthm}
\newtheorem{theorem}{Theorem}[section]

\theoremstyle{definition}
\newtheorem{definition}[theorem]{Definition}

\newcommand*\emptycirc[1][1ex]{\tikz\draw[thick] (0,0) circle (#1);} 
\newcommand*\halfcirc[1][1ex]{%
    \begin{tikzpicture}
    \draw[fill] (0,0)-- (90:#1) arc (90:270:#1) -- cycle ;
    \draw[thick] (0,0) circle (#1);
    \end{tikzpicture}}

\newcommand*\quartercirc[1][1ex]{%
    \rotatebox{90}{\begin{tikzpicture}
    \draw[fill] (0,0) -- (90:#1) arc (90:0:#1) -- cycle ;
    \draw[thick] (0,0) circle (#1);
    \end{tikzpicture}}}
\newcommand*\fullcirc[1][1ex]{\tikz\fill (0,0) circle (#1);}

\newcommand{\rot}[1]{\rotatebox{90}{#1}}

\begin{document}

\title{SoK: Content Moderation for End-to-End Encryption}

\author{Sarah Scheffler}
\affiliation{%
\institution{Princeton University}
\country{USA}
}
\author{Jonathan Mayer}
\affiliation{%
\institution{Princeton University}
\country{USA}
}

\begin{abstract}
Popular messaging applications now enable end-to-end-encryption (E2EE) by default, and E2EE data storage is becoming common. These important advances for security and privacy create new content moderation challenges for online services, because services can no longer directly access plaintext content.

While ongoing public policy debates about E2EE and content moderation in the United States and European Union emphasize child sexual abuse material and misinformation in messaging and storage, we identify and synthesize a wealth of scholarship that goes far beyond those topics. We bridge literature that is diverse in both content moderation subject matter, such as malware, spam, hate speech, terrorist content, and enterprise policy compliance, as well as intended deployments, including not only privacy-preserving content moderation for messaging, email, and cloud storage, but also private introspection of encrypted web traffic by middleboxes. 

In this work, we systematize the study of content moderation in E2EE settings. We set out a process pipeline for content moderation, drawing on a broad interdisciplinary literature that is not specific to E2EE.
We examine cryptography and policy design choices at all stages of this pipeline, and we suggest areas of future research to fill gaps in literature and better understand possible paths forward.

\end{abstract}


\maketitle

\section{Introduction}
\label{sec:intro}

How can an online service implement content moderation when it cannot access content? This challenge, at the intersection of promoting information security and mitigating societal harms, is now a global public policy flashpoint and a focal point for scholarship.

Encryption has seen widespread adoption over the past decade and provides a core component of user security and privacy online.
End-to-end encryption (E2EE) is available and often the default in popular messaging applications~\cite{signal,whatsapp,imessage,line,googleMessages,skypePrivate}, 
and it is the norm for web traffic between clients and servers~\cite{TLS,ipsec}. It is also available in email~\cite{openpgp}, file storage and sharing~\cite{appleE2eeStorage,keybase}, and audio/video chat~\cite{zoom,jitsi,facetime}. Encryption of data at rest, including full-disk and file-based encryption, is common for devices~\cite{bitlocker,veracrypt,truecrypt,macosfde,luks,iosfde,androidfbe}. 
In E2EE, a service platform cannot read or tamper with users' plaintext content. This provides privacy and security not only against external attackers but also against threats that compromise the platform, including data breaches, malicious insiders, and more problematic spying by the platform itself \cite{unger2015sok,signal}.
The security and privacy provided by E2EE advances safeguard users worldwide, protecting journalists, activists, government officials, business leaders, and ordinary users alike.

However, E2EE provides this privacy to both use and abuse. Without access to plaintext or an ability to decrypt, an online service is limited in how it can respond to harmful content and facilitate accountability for criminal acts.

In a 2014 address, then-FBI Director James Comey characterized the challenge for law enforcement as ``going dark''~\cite{goingDark}.  Investigators were struggling to conduct electronic surveillance and obtain electronic evidence, because they were increasingly encountering encryption that online services and device vendors could not bypass.
Comey's remarks tapped into an existing debate \cite{landau2011,bellovin2012going,keysunderdoormats} and set the stage for another decade of encryption policy tussles, initially centered on law enforcement access to data stored on devices  and more recently emphasizing child exploitation that uses E2EE messaging and storage~\cite{goingDark,kardefelt2020encryption,ncmecE2E,internationalStatement,noplacetohide}.

As child exploitation activities moved online, efforts to identify and investigate those activities also moved online.
One of the primary techniques for proactive detection of online child abuse was---and remains---hash matching. In these detection systems, an online service hashes user content and compares the value against a database of known hashes of child sexual abuse material (CSAM). 
These hash databases are often coordinated by national child safety organizations, such as the National Center for Missing and Exploited Children (NCMEC) in the United States. NCMEC began assembling hashes of CSAM in the early 2000s, and it accelerated the practice in 2008 after adopting the PhotoDNA perceptual hash function---which is still in widespread use.
Hash matching remains a best practice in CSAM detection \cite{kardefelt2020encryption,saferio,googlehash,cybertiplinedata,national2016,iwfhash},
and it has also been used or proposed to detect terrorist content \cite{gifct}, misinformation \cite{reis2020detecting}, and suspicious web links \cite{safebrowsing}.

Hash-based detection methods, and more generally content-based detection methods, are not directly implementable in an E2EE setting because the online service cannot analyze content. In response, law enforcement agencies and child safety groups called for a halt to E2EE adoption until similar methods of detecting CSAM and other forms of online child abuse were available~\cite{ncmecE2E,noplacetohide,internationalStatement}.

The encryption policy debate shifted again in 2021, when two independent groups---one in academia and one at Apple---proposed cryptographic protocols that would selectively report user media that matched a perceptual hash set~\cite{USENIX:KulMay21,bhowmick2021apple}. 
If deployed in an E2EE setting, these protocols would essentially create an exception to E2EE for matching content. While some stakeholders applauded these protocols as a breakthrough \cite{thorn2021,wired2021apple}, others (including the academic authors and eventually Apple \cite{wapo,brewster2021apple,wired2022apple}) were more reluctant. The proposals raised more questions than they answered, posing risks to privacy, security, and free expression~\cite{biop,cdtLetter,aclu,eff2021,vergeApple}. Civil society groups were especially alarmed about the possibility that these systems would  quickly expand to categories of content beyond CSAM, especially under pressure from foreign governments~\cite{biop,applePrivacyLetter,cdtLetter,aclu}.

Against this backdrop, security researchers who both appreciate the benefits of E2EE and are concerned about the societal harms that it could facilitate may be left wondering: now what?

This paper seeks to place debates about content moderation in E2EE settings on much-needed shared scientific footing.
In the spirit of earnest intellectual investigation into this divisive topic, we aim to provide an evenhanded systematization of prior work on E2EE content moderation and offer guidance about possible constructive directions for future research.
We go beyond the (important) problem of detecting CSAM and unite diverse areas of literature that address content moderation for E2EE systems, broadly conceived. Our synthesis of relevant prior work spans topics such as preventing spam, ensuring compliance for corporate networks, and defeating malware.
Each problem area poses a distinct set of technical and policy challenges, and while many of these challenges cannot be addressed by technology alone, 
research \textit{can} improve the security, privacy, efficacy, efficiency, and especially transparency of content moderation.

Our aim is that this paper will offer helpful guidance to three interrelated audiences: (1) researchers studying content moderation who wish to learn about possible system designs and open challenges under E2EE; (2) cryptographers and other researchers studying privacy-preserving systems who seek to understand content moderation objectives which might be met through novel designs; and (3) a broad range of stakeholders who are invested in encryption policy and wish to understand the capabilities and limitations of proposed systems in the research literature.

While this paper describes a number of content moderation systems for E2EE, we do not endorse the adoption of any particular design. We are especially concerned about systems that would disclose user content under E2EE, which we do not believe presently offer sufficient assurances of trustworthiness for deployment.

The core of our systematization is organized around a four-part model of content moderation, which we derive from prior work outside the E2EE setting.
This model begins with a \emph{problem context}: the societal harm to be addressed, the role of E2EE in facilitating the harm, the type of content to be moderated, and the parties of concern in sending or receiving the content. The next step is \emph{detection} of the content to moderate, followed by a \emph{response} which may or may not reveal the detection to a third party, such as the online service. Finally, \emph{transparency} enables users to verify and contest the moderation system.
The paper is organized as follows:
\begin{itemize}
\item \textbf{Section \ref{sec:related}} offers 
background on E2EE, content moderation, and the challenges of content moderation under E2EE.
\item \textbf{Section \ref{sec:litsearch}} describes the methods and results of our literature search, with the exception of work on middleboxes (see Appendix~\ref{sec:big-tables}). 
The search involved over 5,000 papers, and we ultimately identified 119 for detailed analysis and synthesis.
\item \textbf{Section \ref{sec:problem-context}} discusses \emph{problem context}. We show how content moderation objectives influence detection and response methods, and we propose future interdisciplinary research to better understand how E2EE interacts with societal harms and how possible content moderation systems could help.
\item \textbf{Section \ref{sec:detection}} explains \emph{detection} paradigms that appear in the literature. We encourage future work evaluating their comparative efficacy and improving perceptual hash functions.
\item \textbf{Section \ref{sec:response}} describes \emph{responses}, some of which implicate user security and privacy. We identify and explain several response mechanisms that are tailored for E2EE settings.
\item \textbf{Section \ref{sec:transparency}} discusses proposed methods for \emph{transparency} in E2EE content moderation systems. We suggest that this is a particularly promising direction for future research. 
\end{itemize}
We hope to encourage future research that explores the design space for content moderation under E2EE, while respecting security, privacy, and accountability.

\section{Background and definitions}
\label{sec:related}

The topic of content moderation under E2EE implicates a vast array of technical and policy literature. Before turning to our survey of prior work, we begin with background on E2EE, content moderation, and the challenges of content moderation under E2EE.

\subsection{End-to-end encryption}
\label{sec:e2ee}
\begin{figure}[ht]
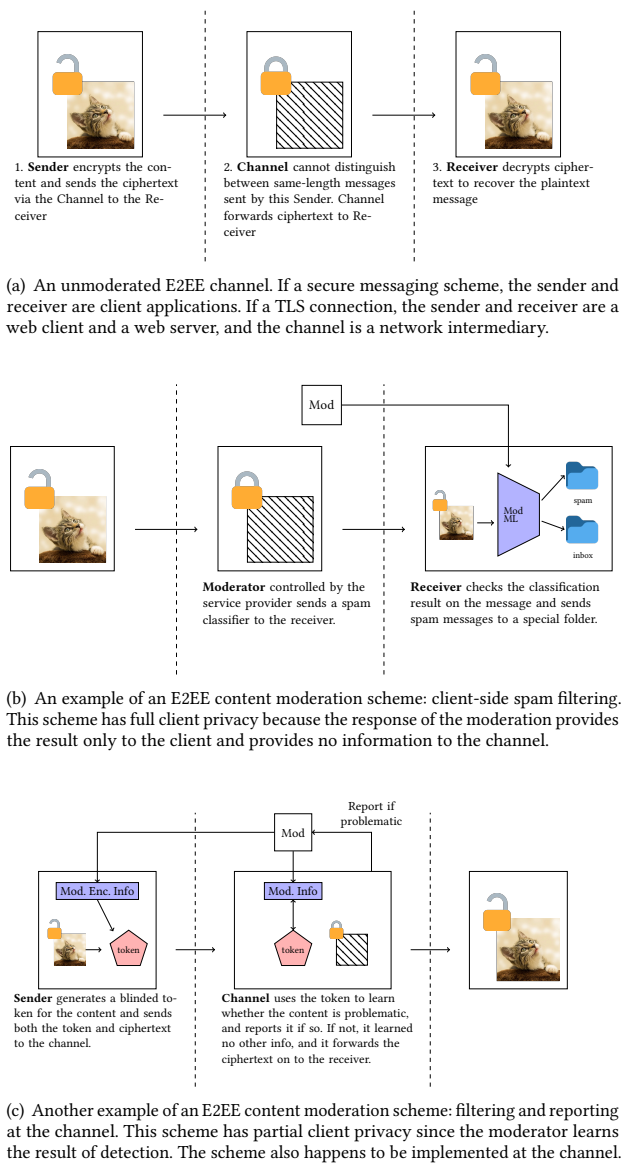

\begin{tabular}{p{0.96\linewidth}}
\subfigure[An unmoderated E2EE channel.  If a secure messaging scheme, the sender and receiver are client applications.  If a TLS connection, the sender and receiver are a web client and a web server, and the channel is a network intermediary.] {%
\resizebox{\linewidth}{!}{
\input{img/e2ee-channel}
\label{fig:e2ee-unmoderated}
}} \\
\subfigure[An example of an E2EE content moderation scheme: client-side spam filtering.  This scheme has full client privacy because the response of the moderation provides the result only to the client and provides no information to the channel.] {%
\resizebox{\linewidth}{!}{
\input{img/example1}
}} \\
\subfigure[Another example of an E2EE content moderation scheme: filtering and reporting at the channel. This scheme has partial client privacy since the moderator learns the result of detection. The scheme also happens to be implemented at the channel.] {%
\resizebox{\linewidth}{!}{
\input{img/example2}
}}
\end{tabular}
\caption{The model of our underlying E2EE communication channel, and two examples of moderation schemes.}
\label{fig:comms-model}
\end{figure}

\emph{End-to-end encryption} (E2EE) refers to any authenticated encryption scheme where the ``ends'' of the communication (a ``sender'' and one or more ``receivers'') can send messages to each other via an abstract central channel and where the channel does not have the cryptographic material necessary to read or invisibly alter the message (see Figure \ref{fig:e2ee-unmoderated}). This model captures E2EE in one-to-one communication (one receiver), group communication (many receivers), and online storage (the receiver is also the sender). We formally define an E2EE scheme as follows.\footnote{For a more detailed description of security properties for E2EE messaging, see  Unger et al.\ \cite{unger2015sok}. Table II in that work shows near-universal agreement on confidentiality, integrity, and authentication, but some disagreement on other properties.}

\begin{definition}[End-to-end encryption]
Communication between at least two client ``ends'' (a sender and one or more receivers) over a channel is \emph{end-to-end encrypted} if it has the following properties:
\begin{enumerate}
\item \emph{Confidentiality}: The plaintext content of the message is indistinguishable from random under chosen plaintext attack by both network attackers and the operator of the channel facilitating message transmission.
\item \emph{Integrity}: The receiver of a message can tell if a received message, along with associated header information, was modified from the sender's original message.
\item \emph{Authentication}: The ends of the communication can confirm each other's identities with long-term cryptographic secrets.
\end{enumerate}
\end{definition}
These properties protect communication under Authenticated Encryption with Associated Data (AEAD)\ \cite{CCS:Rogaway02}, providing IND\$-CPA confidentiality\ \cite{CCS:Rogaway02} for the message, integrity for the message and a public header, and authentication for each end's identity.

Recent debates about content moderation for E2EE have predominantly focused on secure messaging applications, because of their growing popularity. There has also been a convergence in the technical implementation of E2EE for messaging: the
Signal double-ratchet protocol~\cite{signal}, a successor to the Off-the-Record protocol~\cite{OTR}, has been implemented with slight variations by messaging and audio/video chat services~\cite{signal,whatsapp,googleMessages,skypePrivate,facebookMessengerE2EE,jitsi}.

There are other applications and implementations of E2EE, however, that feature much less prominently in encryption policy tussles and that offer lessons for content moderation. 
For example, an HTTPS network connection encrypted with Transport Layer Security (TLS~\cite{TLS}) is end-to-end encrypted from the perspective of a network intermediary, such as a network operator or a security service provider monitoring traffic. In this scenario, the web client and web server are the ends of E2EE communication and the network intermediary is part of the channel. While some methods for TLS content moderation sacrifice security and privacy by terminating the connection at a middlebox with plaintext access~\cite{jarmoc2012ssl,de2020survey},
there has been substantial research on alternative approaches. Some designs use machine learning to analyze encrypted traffic flows~\cite{aldweesh2020deep,sikos2020packet,lee2021towards,keshkeh2021review}), and others use cryptographic methods to perform deep packet inspection (see Appendix~\ref{sec:big-tables}).
These methods are applicable to content moderation goals in school or home networks (e.g., \cite{ramezanian2021parental}) in addition to enterprise policy compliance. We include all of these settings and more in our analysis of content moderation under E2EE.

Signal and the latest version of TLS
(1.3) also both provide two additional properties \cite{canetti2022universally,TLS,unger2015sok}:
\begin{enumerate}
\setcounter{enumi}{3}
\item \emph{Forward secrecy}: Even if the ends' current keys are compromised, messages sent in the past remain confidential.
\item \emph{Deniability}: No coalition of parties, including the receiver of a message, can prove a sender sent a specific message.
\end{enumerate}
These properties can both be considered anti-surveillance measures:
forward secrecy prevents, for example, routers from archiving encrypted traffic and then decrypting later after a key breach.
Deniability ensures that, cryptographically speaking, there is no way to prove that a particular message was sent by a particular sender.\footnote{Deniability occasionally uses different meanings in different areas of literature; here we refer to it to mean ``message repudiation'' as defined by Unger et al.\ \cite{unger2015sok}.}

Signal additionally provides the property of \emph{post-compromise security}:
after a key compromise, the clients can go through a ``refresh'' protocol that ensures future messages they send cannot be read using the compromised keys \cite{canetti2022universally}.

In this paper we are primarily concerned with non-anonymous E2EE, where the central server knows the identity of the sender and receiver(s), because that is the typical design of widely deployed E2EE services at present. There are ``metadata-private'' E2EE messaging schemes that provide anonymity for senders~\cite{signalSealedSender,tor,mixminion}, receivers~\cite{broadcast}, or both~\cite{torhiddenservices,dissent,verdict,ricochet}.

\subsection{Content moderation}

Online services and network operators have devised a vast and diverse toolkit for addressing problematic content. 
Prior work on content moderation is extensive and includes taxonomies of content moderation remedies and approaches~\cite{goldman2021,oecdReport}, analysis of user behavior~\cite{kollock1996managing,seo2019trust}, examination of human and technical challenges with current content moderation systems~\cite{gillespie2020content,gorwa2020algorithmic,DBLP:journals/tochi/JhaverGBG18,matias2019civic,roberts, seering2020reconsidering,gillespie2020expanding,halevy2022preserving,strictercontentmoderation,gillespie,citron2014hate,kaye2019speech},~ and analysis of regulation and governance models~\cite{brown2020regulatory,klonick2017new,grimmelmann2015virtues,sternberg2012misbehavior}, among other topics.  
There is also expansive literature on the problem of automatically detecting particular kinds of content that could merit moderation, such as nudity \cite{ries2014survey,jiao2001detecting,deselaers2008bag,kelly2008screening} and hate speech \cite{fortuna2018survey,aluru2020deep,macavaney2019hate,thomas2021sok,warner2012detecting}.

There are many motivations for content moderation, 
from friendly relocation of off-topic material in a group conversation~\cite{ren2011simulation,offtopicConversation} to removal and reporting of illegal child abuse material~\cite{levyRobinson,appleCommunicationSafety}.
We use the term \emph{problematic content} as an umbrella for user content that an online service may wish to take action on. 

The service may seek to detect problematic content as part of a content moderation scheme, such as by placing the content in a matching dataset or training a machine learning classifier. The content moderation scheme could lead to false positive matches and reveal information about non-problematic content.
Later on, we will refer to this class of problematic content as class $C$ (e.g., the positive results of a spam message classifier) which is attempting to implement class $\tilde{C}$ (e.g., true spam messages).

Several formalized models for content moderation exist in the literature ~\cite{common2020fear,singhal2022sok,cdt}.
We adapt the three-part model of Singhal et al.\ \cite{singhal2022sok} into a four-part model, with the following primary differences: (1) we expand the \emph{terms of use} component into a broader \emph{problem context} component; (2) we expand \emph{enforcement} into \emph{response}, because strict enforcement is just one of many possible responses; and (3) we add a \emph{transparency} component that encompasses how users can verify that content moderation systems are functioning as described.
Our content moderation framework is as follows:
\begin{enumerate}
\item \emph{Problem context}: What is the goal of the content moderation? In particular, what is the societal harm to address, how does E2EE relate, who are the parties of concern, and what is the definition of relevant content? The problem context scopes the other three phases, placing limits on the detection, response, and transparency methods that would be viable.
\item \emph{Detection}: What method is used to identify the content for moderation, and how accurate is it? What privacy protections are used to process the message content within E2EE? Where will the detection be performed---on the sender device, by the channel, on a receiver device, or some combination?
\item \emph{Response}: When a detection occurs, what happens? Which parties are automatically informed, and what additional information is sent to those parties?  What actions are taken automatically? What manual actions are made available?
\item \emph{Transparency}: What information is disclosed to users about the system's purpose, methods, and effectiveness? How can users verify that the system is functioning as described? How can users contest content moderation actions?
\end{enumerate}

These four phases represent four fundamental choices that must be made in the design space of a content moderation system under E2EE. The relationship between these choices can be nuanced. Selecting a machine learning approach to detection instead of matching against a dataset, for example, may reduce confidence in content identification, therefore a weaker response or heightened transparency may be appropriate. 

After describing our literature search in Section~\ref{sec:litsearch}, the paper is organized by these content moderation phases. For each component of a content moderation system, we characterize current research and recommend future directions.

\subsection{Content moderation under E2EE}

The challenges of content moderation for E2EE predominantly relate to maintaining data confidentiality for servers and clients. We discuss these considerations in turn, followed by particular risks of content moderation under E2EE and limitations of this work.

\paragraph{Server privacy}
Content moderation may make use of secret information held by a service provider or a third party. Systems for detecting CSAM, for example, typically rely on matching against a sensitive dataset of CSAM hashes that is kept secret to protect investigative methods and prevent evasion that could reveal investigative methods~\cite{saferio,googlehash,bhowmick2021apple,USENIX:KulMay21}. When the service provider has secret information that must be kept from the 
user during content moderation, we refer to this property as \emph{server privacy}~\cite{USENIX:KulMay21}.
\begin{definition}[Server privacy]
A content moderation scheme has \emph{server privacy} if it maintains the confidentiality of service provider or third-party secrets that are used in the scheme.
These secrets could involve, for example, hashes of known harmful content or a fragile machine learning classifier.
Formally, a content moderation scheme has server privacy if a computationally bounded client has at most negligible advantage in a security parameter at determining whether they are interacting with the real content moderation scheme as opposed to an ``ideal'' scheme where a corrupt client learns only the responses to chosen content.
\end{definition}

While the server privacy property is not unique to E2EE settings, it poses a significant challenge under E2EE because a service provider cannot trivially implement all content moderation server-side. Server privacy also poses significant difficulty for transparency, because a client cannot readily verify that the server is only detecting the content it claims it is detecting.  In Section \ref{sec:transparency}, we discuss recent proposals for verifying important properties of server secrets and suggest directions for future work on the topic.

\paragraph{Client privacy}

The primary concern in content moderation under E2EE, in comparison to ordinary content moderation, is respecting~\emph{client privacy}. E2EE establishes full confidentiality for content against the service provider and other third parties to a communication. Any content moderation scheme that automatically discloses information about content to a third party represents a reduction in the fundamental security and privacy guarantee.

If a content moderation system achieves the same confidentiality, integrity, and authenticity guarantees as the underlying E2EE channel, we call it \emph{fully client private}.
\begin{definition}[Full client privacy]
A content moderation scheme over an E2EE channel has \emph{full client privacy} if it maintains the same confidentiality, integrity, and authenticity properties as the underlying E2EE channel.  In particular, both detection and response are conducted in an end-to-end way, without giving any new information to the service provider or another third party.
\label{def:full-client-privacy}
\end{definition}

In some problem contexts, stakeholders may believe that full client privacy does not achieve content moderation goals. Proposals to counter CSAM, for example, often center on alerting child safety groups and law enforcement. System designs like these, which would automatically notify a third party about problematic content, do not offer full client privacy. They could, however, offer a reduced guarantee which we term \emph{partial client privacy}.
\begin{definition}[Partial client privacy]
A content moderation scheme over an E2EE channel has \emph{partial client privacy with respect to class $C$} if it maintains the same confidentiality, integrity, and authenticity properties as the underlying E2EE channel for all messages \emph{except} those in class $C$. For messages in class $C$ the scheme maintains the integrity and authenticity guarantees but may not provide confidentiality against designated third parties. 
\label{def:partial-client-privacy}
\end{definition}

The class $C$ of messages for which the confidentiality guarantee does not hold could be positives of a ML classification scheme, items that share a perceptual hash with a list, or other categories of content.
The ideal version of this class, $\tilde{C}$, is a theoretical class (like ``content perceptually similar to items on a particular list'') that is measured imperfectly by the real class $C$ (e.g., ``content sharing a PDQ hash with an item on that list'').  This measurement will have both false positives and false negatives, and
the false positive rate 
is the frequency of non-problematic messages whose confidentiality was nevertheless breached by the content moderation system.

In partially client-private systems, the moderator is in essence granted a special key that can be used to read elements of $C$ as an exception to the E2EE system.
This makes the choice of ``ideal'' class $\tilde{C}$ and trustworthiness of implementation $C$ of utmost importance:
an arbitrary or corrupted $C$ could effectively reduce the entire channel to an unencrypted one.
Although a full policy treatment is out of scope of this paper, we discuss some approaches to proving information about $C$ in Section \ref{sec:transparency}.

\paragraph{The terminology of full and partial client privacy}
The term ``client privacy'' was originally used by Kulshrestha and Mayer \cite{USENIX:KulMay21} to mean what we here call ``partial client privacy.''
We use the terms \emph{full client privacy} and \emph{partial client privacy} for two purposes.
First, we seek to emphasize that a scheme that does not achieve even partial client privacy should not be called E2EE at all.
Second, there is a meaningful difference in the privacy guarantee offered by full and partial client privacy:
the channel or moderator effectively has keys for content in $C$, which makes the choice and trustworthiness of $C$ of utmost importance.
One could argue that partially client private systems, too, should not be called E2EE. Some civil society groups and researchers take this position~\cite{cdt,eff2021,eff2019}.
While these observers are very uneasy with partial client privacy, and we respect their perspective, we consider partial client privacy within the scope of this systematization. The concept, as we formalize it in Definition~\ref{def:partial-client-privacy}, is self-consistent and central to the current discourse on content moderation and E2EE (e.g.,\ \cite{biop,USENIX:KulMay21,bhowmick2021apple,levyRobinson,cdt}). 
We hope the term ``partial client privacy'' makes clear that the concept has a coherent definition, that it still provides meaningful security and privacy guarantees, and that it represents a significant departure from the typical E2EE setting of full client privacy.

\paragraph{Maintaining indistinguishability for non-problematic content}
Note that our conception of E2EE-compatible content moderation maintains an \emph{indistinguishability} notion of confidentiality for content outside class $C$. Formally, we require at least indistinguishability against chosen plaintext attack~\cite{katzLindell} for this content.
We therefore exclude moderation that functions by, for example, sending hashes of all messages to the server to perform a match.  
Such an approach would allow a service to check whether a message contained any particular piece of media (say, a divisive political meme) by hashing the media and comparing the hash of the message.
The service provider could essentially monitor user content, well beyond class $C$. We believe these constructions so completely defeat E2EE guarantees that they cannot defensibly be considered compatible, and we emphatically reject these proposed directions from both researchers and governments 
(e.g.,\ \cite{agarwal2021india,DBLP:conf/cvpr/SinghF19,kim2020fast}).

\paragraph{The risks of content moderation under encryption} 
\label{sec:surveil}
The implementation of any content moderation system under E2EE reduces barriers to future surveillance.
Service providers have mixed records responding to external pressure to monitor or censor content \cite{appleNavalny,appleNYT,appleFBI}.
Due to the added power of the server to read some messages in partially client private systems, we see partial client privacy as an especially vulnerable setup: there will be more pressure to monitor more kinds of content for a variety of purposes, and the system will present a more attractive target for external attackers wishing to exploit the system, complicating one of the core benefits of E2EE.
The risks are lower, but not zero, for fully client private systems: these can be adapted into partially client private systems by changing a small amount of client code to report detections to the server rather than keeping them on the client device \cite{biop}.

Additionally, it is a well-documented phenomenon that even if a particular deployer of a content moderation scheme keeps the system tailored for a narrow purpose, other organizations may reuse the same system in more censorious settings
\cite{raman2020measuring,opennet2011west}. 

These topics emerge in many areas of tension between law enforcement, safety, and privacy; prior work offers thorough descriptions of the risks and mechanisms of bypassing encryption to expand surveillance  \cite{carnegiePrinciples,nationalacademies,keysunderdoormats,biop} or censorship \cite{bloch2021content,brown2020regulatory,strictercontentmoderation}.
Scholarship appropriately takes these risks very seriously. We do, however, hold out hope that further research in this field will improve the frontier of possible tradeoffs  and could lead to systems that improve content moderation while maintaining strong security and privacy.

\paragraph{Limitations}
\label{sec:limitations}

Our work has three main limitations.
First, we focus on content moderation performed by a centralized service.
While we touch on user-driven content moderation methods, we do not explore the broad space of designs that could integrate community decision making into E2EE. We believe many forms of collective and delegated user-driven content moderation are feasible under E2EE, by implementing threshold and permission properties within cryptographic protocols. These constructions would maintain full client privacy and are a fruitful direction for future work.

Second, the focus of this work is on content moderation ``of content'' as opposed to other forms of moderation like blocking particular users, building user reputation, or verifying the identities of senders.
A wide literature on these topics exists even in anonymous settings
\cite{SP:RosMalMie22,EPRINT:DHSZZ21,EPRINT:YALXY17,ACNS:BCKSD14,CCS:AuKap12,DBLP:conf/icpads/XuAG12,SP:HenGol11a,DBLP:conf/etrics/KopsellWF06}.
We focus on moderation of content rather than users because it has been the most contested territory for E2EE moderation.

Finally, since we included fully client-side moderation approaches in our literature search, we restricted the search to papers which mentioned encryption or privacy explicitly.
A full literature search of text and image classification methods for content moderation is beyond the scope of this work.

\section{Literature search: methods for content moderation under E2EE}
\label{sec:litsearch}
\label{sec:lit-search}
\label{sec:current}

In this section we describe we describe our literature search and its initial findings.

\begin{table*}[ht]
\resizebox{\linewidth}{!}{%
\begin{tabular}{p{0.15\linewidth}|p{0.5\linewidth}|p{0.24\linewidth}|p{0.26\linewidth}}
\textbf{Moderation Goal} & \textbf{Summary of Archetype} & \textbf{Sub-Archetypes} &  \textbf{Works} \\ \hline \hline
\multirow{3}{=}{Corporate network monitoring} & 
\multirow{3}{=}{A ``middlebox'' may act as a firewall, aim to detect intrusions, malicious data exfiltration, or act as some other policy-based content blocker.  
Generally aims for partial client privacy.
Often has server privacy to increase difficulty of evasion and protect intellectual property.
See Appendix \ref{sec:big-tables} for more details about this setting.} & 
      MPC or Searchable Encryption & 
      \cite{USENIX:GAZBW22,jia2022encrypted,chen2022privacy,li2022towards,ASIACCS:KCBSPN21,poh2021survey,lai2021practical,canard2021towards,ren2020privacy,ESORICS:NHPXLWD20,fan2020group,bkakria2020privacy,CCS:NPLCC19,ren2019toward,desmoulins2018pattern,guo2018enablingA,alabdulatif2017privacy,canard2017blindids,fan2017spabox,li2017clouddpi,yuan2016privacy,melis2016private,lan2016embark,lin2016privacy,sherry2015blindbox} 
      (total: 25) \\ \cline{3-4}
& & Trusted Execution Environment & \cite{yao2022privacy,nikbakht2022chuchotage,poh2021survey,han2020secure,CCS:DWYZWR19,wang2019tvids,trach2018shieldbox,kuvaiskii2018snort,poddar2018safebricks,han2017sgx,trach2017slick,coughlin2017trusted,schiff2016pri,shih2016s} 
(total: 14) \\ \cline{3-4}
& & Other & \cite{ren2021enabling,poh2021survey,goltzsche2018endbox,shi2015privacy,zhou2015towards} (total: 5) \\ 
\hline \hline
\multirow{3}{=}{User reporting (UR) of harassment, abuse, etc.} &
\multirow{3}{=}{In secure messaging, enable users to report abusive or misleading messages to a moderator.  Message franking (see Section \ref{sec:message-franking}) introduces additional integrity guarantees.} &
    Message franking & \cite{USENIX:IssaAlHVar22,jiang2022report,yamamuro2021forward,hirose2020compactly,C:TGLMR19,C:DGRW18,EPRINT:CheTan18,EPRINT:LeoVau18,EPRINT:HugLeo18,C:GruLuRis17} (total: 10) \\ \cline{3-4}
& & Reveal source, traceback, or popular messages & \cite{USENIX:IssaAlHVar22,facts,CCS:PeaEskBon21,CCS:TyaMieRis19} (total: 4) \\ \cline{3-4}
& & Other user reporting & \cite{kazemi2021tiplines,martins2021fact,bagade2020kauwa,meedan2020one,melo2019whatsapp,ghatte2017study,zhang2013safe,lahmadi2011hinky,yan2006enhancing,kong2005scalable,damiani2004p2p} (total: 11) \\ 
\hline \hline
\multirow{4}{=}{Spam filtering} &
\multirow{4}{=}{In secure messaging or E2EE email, prevent high-volume spam, especially those containing scams.  Full client privacy achieved, may or may not have server privacy.} &
    AI/ML via general crypto or MPC & \cite{EPRINT:RRDNA21,wang2020privacy,bian2019towards,ryffel2019partially,gupta2017pretzel,costantino2017privacy,khedr2015shield,zhang2015pif,pathak2011privacy} (total: 9) \\ \cline{3-4}
& & AI/ML or matching fully client-side & \cite{googleSpamProtection,agarwal2022jettisoning,ghatte2017study,verma2012detecting,taufiq2012simple,lahmadi2011hinky,yan2006enhancing,kong2005scalable,damiani2004p2p} (total: 9) \\ \cline{3-4}
& & Metadata-based & \cite{nabeel2021cadue,cashweb,zhang2013safe,wang2007using,EPRINT:JakLinAlg03} (total: 5) \\ \cline{3-4}
& & Other & \cite{PKC:NSSWXZ21,tarafdar2021spam,schlogl2020ennclave} (total: 3) \\
\hline \hline
\multirow{4}{=}{Malware/phishing, ``safe browsing''} &
\multirow{4}{=}{In web browsing, messaging, or E2EE file transfer/storage, detect if a particular file or URL is suspicious or malware.  Typically has full client privacy, may or may not have server privacy.
Omitting 155 papers for detecting malware in encrypted TLS traffic by performing ML classification on the encrypted traffic flow; see surveys \cite{aldweesh2020deep,sikos2020packet,lee2021towards,keshkeh2021review,shbair2020survey}.} &
    Matching via general crypto or MPC & \cite{USENIX:KogCor21,EPRINT:SCDGGY21,ramezanian2020private,cui2019ppsb,hwang2019static,cui2018towards,sun2017primal,poon2016scanning}
    (total: 9) \\ \cline{3-4}
& & Client-side or metadata-based & \cite{kucuk2018bigbing,whatsappsuspiciouslinks,verma2012detecting,jammalamadaka2005pvault} (total: 4) \\ \cline{3-4}
& & AI via MPC or federated learning & \cite{EPRINT:SCDGGY21,shaik2021privacy,galvez2020less,chou2020privacy} (total: 4) \\ \cline{3-4}
& & Matching in Trusted Execution Environment & \cite{wei2022epmdroid,deyannis2020trustav,ASIACCS:TLPEPA17} (total: 3) \\
\hline \hline
\multirow{2}{=}{Parental or educational control} & 
\multirow{2}{=}{A typical setting is to detect or block specific keywords, websites, or content in TLS traffic, usually with no special hardware.} & 
      MPC or Searchable Encryption & \cite{USENIX:GAZBW22,jia2022encrypted,ramezanian2021parental,poh2021survey,lai2021practical,canard2021towards,fan2020group,CCS:NPLCC19,ramezanian2019privacy,canard2017blindids,li2017clouddpi,lan2016embark,sherry2015blindbox} (total: 13) \\ \cline{3-4}
& &  Trusted Execution Environment & \cite{schiff2016pri} (total: 1) \\
\hline \hline
\multirow{4}{=}{Child safety} &
\multirow{4}{=}{In secure messaging or video chat, detect child sexual abuse.  To detect imagery or video, either match against a list using a PHF, or use ML.  Server privacy generally considered required.} &
  CSAM detection via client-side AI & \cite{appleCommunicationSafety,dragonflai,t3k,galaxkey} (total: 4) \\ \cline{3-4}
& & Matching with a server-held list of CSAM & \cite{bhowmick2021apple,USENIX:KulMay21,cyanProtect} (total: 3) \\ \cline{3-4}
& & CSAM detection via filename metadata & \cite{pereira2020metadata,al2020file,panchenko2012detection} (total: 3) \\ \cline{3-4}
& & Other child safety & \cite{ramezanian2019privacy,safetonet} (total: 2) \\ 
\hline \hline
\multirow{4}{=}{Other} &
\multirow{4}{=}{Papers for moderation of content with few results.  Note that this field contains most of the ``standard'' content moderation topics.} &
    Mis/disinformation & \cite{kazemi2021tiplines,meedan2020one,bagade2020kauwa,reis2020detecting,melo2019whatsapp,freitas2019can} (total: 6) \\ \cline{3-4}
& & Hate/harassment & \cite{martins2021fact,ramezanian2019privacy,reich2019privacy,zhang2013safe} (total: 4)  \\ \cline{3-4}
& & Nudity/NSFW & \cite{pandey2021device,ryffel2019partially} (total: 2) \\ \cline{3-4}
& & Terrorism \& violent extremism & \cite{costantino2017privacy,USENIX:KulMay21} (total: 2) \\ \cline{3-4}
\hline \hline
\end{tabular}
} 
\caption{Literature search results for content moderation under E2EE sorted by goal.  Some works appear in multiple categories.}
\label{tab:litsearch}
\end{table*}

\paragraph{Literature search and query terms}

We provide a short summary of our literature search methods here, and we include full details in Appendix~\ref{app:methods}.
Our literature search initially surfaced papers by running the following queries in August 2022 in computer science and cryptography-related academic venues: content moderation, CSAM, end to end, malware, misinformation, porn, pornography, and spam.
The academic venues were 
ACM CCS,
CRYPTO,
NDSS,
PETS,
IEEE S\&P,
Usenix Security,
arXiv CS,
and
IACR ePrint.
We additionally examined the top 200 results from Google Scholar for ``encrypted content moderation'' and ``end to end encrypted content moderation,'' the five entries to the recent UK Safety Tech Challenge (UK STC \cite{ukSafetyTechWinners}), and the documentation for Apple iMessage, Google Messages, Signal, and WhatsApp.
See Appendix \ref{app:methods} for all exact queries for each venue.
These queries formed the initial set of works.
We manually inspected those papers to identify relevant works using the criteria listed below,
and we then identified additional relevant papers using snowball sampling.

We examined the papers manually to identify both their relevance for inclusion, and to extract the information shown in Tables \ref{tab:individual-nonmiddlebox} and \ref{tab:individual-middlebox}.
To be included for analysis, a work must have had at least one subsection in which it describes, constructs, or implements a content moderation system (thus excluding generic cryptography papers that could be applied to content moderation), and it must achieve at least partial client privacy according to Definition \ref{def:partial-client-privacy}.  Client-side systems must have mentioned that they intended to be used in an encrypted or private setting in order to be included, thus excluding a large number of papers on generic content moderation that could be run on client devices.

\begin{savenotes}  
\begin{table*}
\resizebox{!}{4.18in}{
\begin{tabular}{lll|cccccc|cccc|lll|l|cc|cc|cl}
\multicolumn{3}{c|}{\textbf{\ul{Problem context}}} & \multicolumn{16}{c|}{\textbf{\ul{Detection}}} & \multicolumn{2}{c|}{\textbf{\ul{Response}}} & \multicolumn{2}{c}{\textbf{\ul{Transparency}}}\\
\textbf{Work} & \textbf{Goal} & \textbf{Medium} & \rot{\textbf{User reporting}} & \rot{\textbf{Metadata}} & \rot{\textbf{Exact matching}} & \rot{\textbf{Perceptual matching}} & \rot{\textbf{Rules/patterns}} & \rot{\textbf{Machine learning}} & \rot{\textbf{General Crypto/MPC}} & \rot{\textbf{Hom./Func.\ Encryption}} & \rot{\textbf{Trusted hardware}} & \rot{\textbf{Client-side}} & \rot{\textbf{Size of blocklist/dataset}} & \rot{\textbf{Latency}} & \rot{\textbf{Communication}} & \rot{\textbf{Predictive Performance}} & \rot{\textbf{Security Against Server}} & \rot{\textbf{Security Against Client}} & \rot{\textbf{Server privacy}} & \rot{\textbf{Client privacy}} & \rot{\textbf{Transparency}} & \textbf{Details}\\
\hline
Nguyen et al.\ \cite{PKC:NSSWXZ21} & Spam & Email &   &  &   &   & \fullcirc &   &   &   &   &   &        & $\times$ & $\times$ &        & Mal. & Mal. & \emptycirc &   & \emptycirc &   \\ 
CADUE\ \cite{nabeel2021cadue} & Spam & Email &   & \fullcirc &   &   &   & \fullcirc &   &   &   &   & 587k & $\times$ & $\times$ & FPR: 0.003 &   &   & \emptycirc & \fullcirc &   &   \\ 
Wang et al.\ \cite{wang2020privacy} & Spam & Email &   &  &   &   &   & \fullcirc & \fullcirc & \fullcirc &   &   & 40k & 22s & $\times$ & Acc: 0.96 & S.H. &   & \fullcirc & \fullcirc & \emptycirc &   \\ 
Bian et al.\ \cite{bian2019towards} & Spam & Email &   &  & \fullcirc &   & \fullcirc &   & \fullcirc & \fullcirc &   &   & 33k & 0.5s & $\times$ & Prec: 0.89 & S.H. & S.H. &   & \fullcirc & \emptycirc &   \\ 
Pretzel\ \cite{gupta2017pretzel} & Spam & Email &   &  &   &   &   & \fullcirc & \fullcirc & \fullcirc &   &   & 33k & 1s & $\times$ & Acc: 0.99 & S.H. & S.H. & \emptycirc & \fullcirc & \fullcirc & Concrete action \\ 
Ghatte and Rajmane\ \cite{ghatte2017study} & Spam & Email & \fullcirc &  &   &   &   &   &   &   &   & \fullcirc & $\times$ & $\times$ & $\times$ & $\times$ & $\times$ & $\times$ &   &   &   &   \\ 
SHIELD\ \cite{khedr2015shield} & Spam & Email &   &  &   &   &   & \fullcirc & \fullcirc &   &   &   & $\times$ & 10s & $\times$ &        & S.H. & S.H. &   & \fullcirc & \emptycirc &   \\ 
Pathak et al.\ \cite{pathak2011privacy} & Spam & Email &   &  &   &   &   & \fullcirc & \fullcirc & \fullcirc &   &   & 206k & 41s &  Asymp. & $\times$ & S.H. & S.H. & \fullcirc & \fullcirc & \emptycirc &   \\ 
Wang and Chen\ \cite{wang2007using} & Spam & Email &   & \fullcirc &   &   &   &   &   &   &   & \fullcirc & 11k & $\times$ & $\times$ & Acc: 0.925 &   & S.H. & \emptycirc & \fullcirc & \emptycirc &   \\ 
Yan and Cho\ \cite{yan2006enhancing} & Spam & Email & \fullcirc &  &   &   &   &   &   &   &   & \fullcirc & 200k &  Asymp. & $\times$ & FPR: $10^{-8}$ &   &   &   &   &   &   \\ 
Kong et al.\ \cite{kong2005scalable} & Spam & Email & \fullcirc &  & \fullcirc &   &   &   &   &   &   &   &        &        & $\times$ & $\times$ & S.H. & Mal. &   &   &   &   \\ 
Damiani et al.\ \cite{damiani2004p2p} & Spam & Email & \fullcirc &  & \fullcirc &   &   &   &   &   &   &   & $\times$ & $\times$ & $\times$ &        & $\times$ & S.H. &   & $\star$\footnote{Peers and super-peers learn reported messages.} &   &   \\ 
Jakobsson et al.\ \cite{EPRINT:JakLinAlg03} & Spam & Email &   & \fullcirc &   &   &   &   &   &   &   &   & $\times$ & $\times$ & $\times$ &        &   & Mal. &   &   & \emptycirc &   \\ 
Agarwal et al.\ \cite{agarwal2022jettisoning} & Spam & Messaging &   & \fullcirc &   &   &   & \fullcirc &   &   &   & \fullcirc & 2.6M & $\times$ & $\times$ & Prec: 0.94 &   & Mal. & \emptycirc & \fullcirc &   &   \\ 
Resende et al.\ \cite{EPRINT:RRDNA21} & Spam & Messaging &   &  &   &   &   & \fullcirc & \fullcirc &   &   &   & 5.6k & 0.35s & $\times$ & FPR: 0.179 & S.H. & S.H. & \fullcirc & \fullcirc & \emptycirc &   \\ 
Tarafdar et al.\ \cite{tarafdar2021spam} & Spam & Messaging &   &  &   & \fullcirc &   &   &   &   &   & \fullcirc & 50 & $\times$ & $\times$ & Acc:\footnote{Extremely small sample size.} 1 &   &   & \emptycirc & \fullcirc &   &   \\ 
Google Spam Protection\ \cite{googleSpamProtection} & Spam & Messaging &   &  &   &   &   & \fullcirc &   &   &   & \fullcirc & $\times$ & $\times$ & $\times$ & $\times$ &   &   & \emptycirc & $\star$\footnote{Reveals telephone number.} &   &   \\ 
Nuruzzaman et al.\ \cite{taufiq2012simple} & Spam & Messaging &   &  &   &   &   & \fullcirc &   &   &   & \fullcirc & 875 & 0.5s & $\times$ & Acc: 0.983 &   &   & \emptycirc & \fullcirc &   &   \\ 
Hinky\ \cite{lahmadi2011hinky} & Spam & Messaging & \fullcirc &  & \fullcirc &   &   &   &   &   &   &   & 50M & 0.1s &        & FPR: Param. &   &   & \emptycirc & \fullcirc &   &   \\ 
CashWeb\ \cite{cashweb} & Spam & Other &   & \fullcirc &   &   &   &   &   &   &   &   & $\times$ & $\times$ & $\times$ & $\times$ & S.H. & S.H. &   &   & \emptycirc &   \\ 
Zhang et al.\ \cite{zhang2015pif} & Spam & Other &   & \fullcirc &   &   & \fullcirc &   &   &   &   &   & 128k & $\times$ & $\times$ & $\times$ & S.H. & S.H. & \fullcirc & \halfcirc & \emptycirc &   \\ 
eNNclave\ \cite{schlogl2020ennclave} & Spam & Unspecified &   &  &   &   &   & \fullcirc &   &   & \fullcirc &   & 150k & 100s & $\times$ & Acc: 0.744 &   & S.H. & \fullcirc & \fullcirc & \fullcirc & Concrete action \\ 
Ryffel et al.\ \cite{ryffel2019partially} & Spam, Nudity & Messaging &   &  &   &   &   & \fullcirc & \fullcirc & \fullcirc &   &   & $\times$ & $\times$ & $\times$ & Acc: 0.98 & S.H. & S.H. & \fullcirc & \fullcirc & \emptycirc &   \\ 
Constantino et al.\ \cite{costantino2017privacy} & Spam, TVEC & Messaging &   &  &   &   &   & \fullcirc & \fullcirc & \fullcirc &   &   & 308 & 1147s & $\times$ & $\times$\footnote{Crypto induces negligible additional errors. Accuracy of the classifier not given.\label{foot:cryptoacc}} & S.H. & Mal. & \fullcirc & \halfcirc & \emptycirc &   \\ 
\hline
Kogan and Corrigan-Gibbs\ \cite{USENIX:KogCor21} & Security & Browsing/TLS &   &  & \fullcirc &   &   &   & \fullcirc &   &   &   & 3M & 0.01s  & 1 KB & FPR: Negl. & Mal.(NC) &   & \emptycirc & \fullcirc & \emptycirc &   \\ 
Shah et al.\ \cite{EPRINT:SCDGGY21} & Security & Browsing/TLS &   &  &   &   &   & \fullcirc & \fullcirc &   &   &   & 10k & 3.6s & 1301MB & Prec: 0.95 & S.H. & S.H. & \fullcirc & \fullcirc & \emptycirc &   \\ 
Chou et al.\ \cite{chou2020privacy} & Security & Browsing/TLS &   &  &   &   &   & \fullcirc & \fullcirc & \fullcirc &   &   & 19k & 0.7s & 250 KB & $\times$ & S.H. & S.H. & \fullcirc & \fullcirc & \emptycirc &   \\ 
Ramezanian et al.\ \cite{ramezanian2020private} & Security & Cloud storage &   &  & \fullcirc &   &   &   & \fullcirc &   &   &   & 2M & 1.8s & 24 KB & FPR: $10^{-4}$ & Mal. & Mal. & \fullcirc & \fullcirc & \emptycirc &   \\ 
Hwang and Yoon\ \cite{hwang2019static} & Security & Cloud storage &   &  & \fullcirc &   &   &   & \fullcirc & \fullcirc &   &   & $\times$ & 246s & $\times$ &        & Mal. &   & \fullcirc & \fullcirc & \emptycirc &   \\ 
PriMal\ \cite{sun2017primal} & Security & Cloud storage &   &  & \fullcirc &   &   &   & \fullcirc &   &   &   & 131k & 3s & 2.5 MB & FPR: $10^{-6}$ & S.H. & S.H. & \fullcirc & \fullcirc & \emptycirc &   \\ 
Poon and Miri\ \cite{poon2016scanning} & Security & Cloud storage &   &  & \fullcirc &   &   &   & \fullcirc & \fullcirc &   &   & $\times$ & $\times$ & $\times$ & $\times$ & S.H. & S.H. & \fullcirc & \fullcirc & \emptycirc &   \\ 
EPMDroid\ \cite{wei2022epmdroid} & Security & Other &   &  & \fullcirc &   &   &   &   &   & \fullcirc &   & 600 & 0.4ms &  Asymp. & FPR: 0.010 & TEE &   & \fullcirc & \fullcirc & \halfcirc & Attestation \\ 
Galvez et al.\ \cite{galvez2020less} & Security & Other &   &  &   &   &   & \fullcirc &   &   &   &   & 40k & 13s & $\times$ & F1: 0.959 & S.H. & Mal. & \emptycirc & \fullcirc & \emptycirc &   \\ 
Cui et al.\ \cite{cui2018towards} & Security & Other &   &  & \fullcirc &   &   &   & \fullcirc &   &   &   & 1260 & 0.14ms & 1.33 KB & FPR: Param. & S.H. & S.H. & \fullcirc & \fullcirc & \emptycirc &   \\ 
Pvault\ \cite{jammalamadaka2005pvault} & Security & Other &   &  &   &   & \fullcirc &   &   &   &   & \fullcirc & $\times$ & $\times$ & $\times$ &        & Mal. &   &   & \fullcirc &   &   \\ 
TrustAV\ \cite{deyannis2020trustav} & Security & Unspecified &   &  &   &   &   &   &   &   & \fullcirc &   & 30k & $\times$ & $\times$ & $\times$ & S.H. & S.H. & \emptycirc & \fullcirc & \halfcirc & Attestation \\ 
BigBing\ \cite{kucuk2018bigbing} & Security & Unspecified & \fullcirc &  & \fullcirc &   &   & \fullcirc & \fullcirc &   &   & \fullcirc & 15k & 0.519s & $\times$ & F1: 0.976 & S.H. & Mal. &   &   &   &   \\ 
Tamrakar et al.\ \cite{ASIACCS:TLPEPA17} & Security & Unspecified &   &  & \fullcirc &   &   &   &   &   & \fullcirc &   & 67M & 0.25ms & $\times$ & FPR: 0.0009 & Mal. & S.H. & \fullcirc & \fullcirc & \emptycirc &   \\ 
Shaik et al.\ \cite{shaik2021privacy} & Security & Browsing/TLS &   &  &   &   &   & \fullcirc & \fullcirc & \fullcirc &   &   & 100k & 9.2s & 131.1 KB & Acc: 0.956 & $\times$ & $\times$ & \fullcirc & \fullcirc & \emptycirc &   \\ 
Verma et al.\ \cite{verma2012detecting} & Security, Spam & Email &   &  &   &   &   & \fullcirc &   &   &   & \fullcirc & 3k & $\times$ & $\times$ & FPR: 0.007 &   &   & \emptycirc & \fullcirc &   &   \\ 
WhatsApp Suspicious Messages\ \cite{whatsappsuspiciouslinks} & Security, Spam & Messaging &   &  & \fullcirc &   & \fullcirc &   &   &   &   & \fullcirc & $\times$ & $\times$ &        & $\times$ &   &   & \emptycirc & \fullcirc & \emptycirc &   \\ 
\hline
Jiang et al.\ \cite{jiang2022report} & UR & Messaging & \fullcirc &  & \fullcirc &   &   &   &   &   &   &   &        & 13ms & $\times$ & FPR: Negl. & S.H. & Mal. &   &   & \emptycirc &   \\ 
Yamamuro et al.\ \cite{yamamuro2021forward} & UR & Messaging & \fullcirc &  &   &   &   &   &   &   &   &   &        & $\times$ &  Asymp. &        &   &   &   &   &   &   \\ 
Hirose\ \cite{hirose2020compactly} & UR & Messaging & \fullcirc &  &   &   &   &   &   &   &   &   &        &        &        &        & Mal. & Mal. &   &   &   &   \\ 
Tyagi et al. (B)\ \cite{C:TGLMR19} & UR & Messaging & \fullcirc &  &   &   &   &   &   &   &   &   &        & 7.3ms & 489 B &        & Mal. & Mal. &   &   &   &   \\ 
Huguenin-Dumittan and Leontiadis\ \cite{EPRINT:HugLeo18} & UR & Messaging & \fullcirc &  &   &   &   &   &   &   &   &   &        &        &        &        & Mal. & Mal. &   &   &   &   \\ 
Chen and Tang\ \cite{EPRINT:CheTan18} & UR & Messaging & \fullcirc &  &   &   &   &   &   &   &   &   &        &        &        &        & Mal. & Mal. &   &   &   &   \\ 
Dodis et al.\ \cite{C:DGRW18} & UR & Messaging & \fullcirc &  &   &   &   &   &   &   &   &   &        & $\times$ & $\times$ &        & Mal. & Mal. &   &   &   &   \\ 
Leontiadis and Vaudenay\ \cite{EPRINT:LeoVau18} & UR & Messaging & \fullcirc &  &   &   &   &   &   &   &   &   &        &        &        &        & Mal. & Mal. &   &   &   &   \\ 
Grubbs et al.\ \cite{C:GruLuRis17} & UR & Messaging & \fullcirc &  &   &   &   &   &   &   &   &   &        &        &        &        & Mal. & Mal. &   &   &   &   \\ 
Hecate\ \cite{USENIX:IssaAlHVar22} & UR, Misinfo & Messaging & \fullcirc &  &   &   &   &   &   &   &   &   &        & 37ms & 380 B &        & Mal. & Mal. &   & $\star$\footnote{Reveals the source.\label{foot:reveal-source}} &   &   \\ 
Peale et al.\ \cite{CCS:PeaEskBon21} & UR, Misinfo & Messaging & \fullcirc &  &   &   &   &   &   &   &   &   &        & 0.057ms & 160 B &        &   & Mal. &   & $\star$\footnote{Reveal the source or forwarding tree.} &   &   \\ 
FACTS\ \cite{facts} & UR, Misinfo & Messaging & \fullcirc & \fullcirc &   &   &   &   & \fullcirc &   &   &   & 1M & 98ms & $\times$ & FPR: Param. & S.H. & Mal. &   & $\star$\footnotemark[\getrefnumber{foot:reveal-source}] &   &   \\ 
Tyagi et al. (A)\ \cite{CCS:TyaMieRis19} & UR, Misinfo & Messaging & \fullcirc &  &   &   &   &   &   &   &   &   &        & 8us & 96 B &        &   & Mal. &   & $\star$\footnote{Reveals the forwarding path.} &   &   \\ 
SAFE\ \cite{zhang2013safe} & UR, Spam & Other & \fullcirc &  &   &   & \fullcirc &   &   &   &   & \fullcirc & 128k & 25s & $\times$ & $\times$ &   &   &   &   & \fullcirc & Concrete action \\ 
\hline
Apple PSI\ \cite{bhowmick2021apple} & CSAM & Cloud storage &   &  &   & \fullcirc &   &   & \fullcirc &   &   &   & $\times$ & $\times$ & $\times$ & PHF & Mal. & Mal. & \fullcirc & \halfcirc & \fullcirc & Concrete action \\ 
Pereira et al.\ \cite{pereira2020metadata} & CSAM & Cloud storage &   & \fullcirc &   &   &   &   &   &   &   &   & 73k & $\times$ &        & Prec: 0.938 &   &   &   &   &   &   \\ 
Al Nabki et al.\ \cite{al2020file} & CSAM & Cloud storage &   & \fullcirc &   &   &   &   &   &   &   &   & 65k & 0.06ms &        & Prec: 0.84 &   &   &   &   &   &   \\ 
iCOP\ \cite{panchenko2012detection} & CSAM & Cloud storage &   & \fullcirc &   &   &   &   &   &   &   &   & 106k & $\times$ &        & Acc: 0.970 &   &   &   &   &   &   \\ 
Cyacomb (UK STC)\ \cite{cyanProtect} & CSAM & Messaging &   &  & \fullcirc &   &   &   & \fullcirc &   &   &   & $\times$ & $\times$ & $\times$ & $\times$ & $\times$ & Mal. & \fullcirc & \fullcirc & \quartercirc & Mention \\ 
Galaxkey (UK STC)\ \cite{galaxkey} & CSAM & Messaging &   &  & \fullcirc &   &   & \fullcirc &   &   &   & \fullcirc & $\times$ & $\times$ & $\times$ & $\times$ & $\times$ &   & \emptycirc &   & \halfcirc & Consent \\ 
Apple Communication Safety\ \cite{appleCommunicationSafety} & CSAM & Messaging &   &  &   &   &   & \fullcirc &   &   &   & \fullcirc & $\times$ & $\times$ &        & $\times$ &   &   & \emptycirc & \fullcirc &   &   \\ 
SafeToNet (UK STC)\ \cite{safetonet} & CSAM & Other &   &  &   &   &   & \fullcirc &   &   &   & \fullcirc & $\times$ & $\times$ & $\times$ & $\times$ & $\times$ &   & \emptycirc & \fullcirc & \quartercirc & Mention \\ 
DragonflAI (UK STC)\ \cite{dragonflai} & CSAM & Unspecified &   &  &   &   &   & \fullcirc &   &   &   & \fullcirc & $\times$ & 60ms &        & F1: 0.979\footnote{Obtained from \url{https://www.dragonflai.co/} on 11/23/2022.} &   &   &   &   & \halfcirc & Consent \\ 
T3K Forensics (UK STC)\ \cite{t3k} & CSAM & Unspecified &   &  &   &   &   & \fullcirc &   &   &   & \fullcirc & $\times$ & $\times$ & $\times$ & Prec: 0.88 &   &   & \emptycirc &   & \emptycirc &   \\ 
Kulshrestha and Mayer\ \cite{USENIX:KulMay21} & CSAM, TVEC & Messaging &   &  &   & \fullcirc &   &   & \fullcirc &   &   &   & 16.7M & 10.5s & 395 KB & PHF & Mal. & Mal. & \fullcirc & \halfcirc & \quartercirc & Mention \\ 
\hline
Filho and Shuen\ \cite{martins2021fact} & Misinfo & Messaging & \fullcirc &  &   &   &   &   &   &   &   & \fullcirc &        &        &        &        &   &   &   &   &   &   \\ 
Kazemi and Garimella\ \cite{kazemi2021tiplines} & Misinfo & Messaging & \fullcirc &  &   &   &   &   &   &   &   &   & 977k & $\times$ & $\times$ &        &   &   &   &   &   &   \\ 
Meedan\ \cite{meedan2020one} & Misinfo & Messaging & \fullcirc &  &   &   &   &   &   &   &   &   & 15k & $\times$ & $\times$ &        &   &   &   &   &   &   \\ 
Reis et al.\ \cite{reis2020detecting} & Misinfo & Messaging &   &  &   & \fullcirc &   &   &   &   &   & \fullcirc & 810k & $\times$ &        & $\times$ &   &   & \emptycirc & \fullcirc &   &   \\ 
Kauwa-Katte Fake News\ \cite{bagade2020kauwa} & Misinfo & Messaging & \fullcirc &  &   &   &   &   &   &   &   &   & $\times$ & $\times$ & $\times$ &        &   &   &   &   &   &   \\ 
Melo et al.\ \cite{freitas2019can} & Misinfo & Messaging &   & \fullcirc &   &   &   &   &   &   &   &   & 400k &        &        &        &   &   &   &   & \emptycirc &   \\ 
Whatsapp Monitor\ \cite{melo2019whatsapp} & Misinfo & Messaging & \fullcirc &  &   &   &   &   &   &   &   &   & $\times$ & $\times$ & $\times$ &        &   &   &   &   &   &   \\ 
\hline
Reich et al.\ \cite{reich2019privacy} & Hate/harass & Messaging &   &  &   &   &   & \fullcirc & \fullcirc &   &   &   & 10k & 2.7s & $\times$ & Acc: 0.744 &   &   & \fullcirc & \fullcirc & \emptycirc &   \\ 
Ramezanian and Niemi\ \cite{ramezanian2019privacy} & Hate/harass & Messaging &   &  &   &   & \fullcirc & \fullcirc & \fullcirc &   &   &   & $\times$ & $\times$ & $\times$ & $\times$ & S.H. & S.H. & \fullcirc & \halfcirc & \emptycirc &   \\ 
Pandey et al.\ \cite{pandey2021device} & Nudity & Other &   &  &   &   &   & \fullcirc &   &   &   & \fullcirc & 85k & 85ms &        & Prec: 0.98 &   &   & \emptycirc &   &   &   \\ 
\hline
\hline
\multicolumn{23}{c}{
\begin{tabular}{llll}
\textbf{General}                               
   & \textbf{Efficiency}      
   & \textbf{Predictive Performance} 
   & \textbf{Security \& Client Privacy} 
\\ $\fullcirc$: Property present                  
   & Blocklist/dataset size is for the largest described.          
   & Highest value given among FPR, Acc, Prec, F1, AUC.
   & S.H.: Semi-honest
\\ $\emptycirc$: Property absent                  
   & Latency and communications are for the size of the largest 
   & FPR: False positive rate 
   & Mal.: Malicious
\\ $\times$: Property could not be determined     
   & blocklist/dataset, where provided.  
   & Acc.: Accuracy (total correct over total classifications)
   & NC: Non-collusion assumption
\\ (Blank): Property irrelevant
   & Asymp.: Asymptotic but no concrete efficiency provided
   & Prec.: Precision
   & TEE: Assumption that TEE use is honest and secure
\\ $\star$: Special (see footnote)
    & \quad
    & F1: F1-score 
    & $\fullcirc$: Full client privacy
\\ \quad
    & \quad
    & Param.: Tunable parameter with FPR/FNR tradeoff
    & $\halfcirc$: Partial client privacy
\\ \quad
    & \quad
    & Negl.: Negligible in a security parameter
    & \quad
\end{tabular}
}
\end{tabular}
} 
\caption{\protect Details of non-middlebox methods for E2EE content moderation found in our survey. See Table \ref{tab:individual-middleboxes} for middleboxes.
}
\label{tab:individual-nonmiddleboxes}
\label{tab:individual-nonmiddlebox}
\label{tab:individual-nonmiddle}
\end{table*}

\paragraph{Results}

Our search resulted in 119 relevant papers including those containing novel cryptographic proposals, metadata-based approaches, and client-side approaches, plus an additional 155 papers on malware detection in TLS traffic by performing machine learning on the encrypted network flow.  We also found 19 papers that would have qualified but did not meet our confidentiality guarantee of indistinguishability on non-matches, 22 relevant surveys (of which 12 were about malicious traffic detection), 18 papers about perceptual hash functions, and numerous papers that concerned the topic of content moderation under encryption but contained no method for performing content moderation in E2EE.
The 119 relevant papers are shown by moderation category in Table \ref{tab:litsearch}.
Figure \ref{fig:results-by-year} shows the results by category and year.
Table \ref{tab:individual-nonmiddle} shows the details of all non-middlebox works; the detailed middlebox results are deferred to Table \ref{tab:individual-middleboxes} in Appendix \ref{sec:big-tables}.

For the remainder of this work, we walk through our findings and suggestions in the four parts of our content moderation pipeline.

\section{Problem Context}
\label{sec:problem-context}

In our terminology, the \emph{problem context} is the externally-provided goal of what kind(s) of content the service provider wishes to moderate.
In Table \ref{tab:litsearch} we sorted the works in E2EE content moderation by their moderation context.
The goal of moderation highly influences the design choices for detection, response, and transparency.
For an example set of considerations on how this is true in or out of the encrypted setting, see three short case studies in Appendix \ref{app:case-studies}.
For the rest of this section, we describe our findings on how the problem context influences the E2EE detection and response, and suggest that future security research be more specialized.

\subsection{Problem context affects choice of detection and response mechanisms}

Naively, we would expect different detection methods to be used in different problem contexts.  Our literature search bears this out, as shown in Tables  \ref{tab:individual-nonmiddle} and \ref{tab:individual-middleboxes}.
Overall,
TLS traffic inspection mainly used rule and pattern matching (93\%), identifying
misinformation relied mainly on user reporting (71\%),
detecting malware URLs or binaries mainly used exact matching (58\%),
and other categories were more mixed.
Interestingly, perceptual hash functions were rare, which we discuss further in Section \ref{sec:phf}.

We also saw differences in the client privacy of the scheme based on problem context.
Works detecting threats or annoyances to the user, such as malware or spam, were nearly universally fully client private.
Works focusing on child safety concerns were more mixed: of the eight proposals whose main focus was child safety, two offered partial client privacy \cite{bhowmick2021apple,USENIX:KulMay21}, two offered full client privacy \cite{appleCommunicationSafety,safetonet}, and the remaining four were prototypes agnostic as to the final setting of client privacy \cite{cyanProtect,galaxkey,dragonflai,t3k}.

Although these choices naively make sense, we are not aware of quantitative research that evaluates the difference in effectiveness of these systems across different problem contexts in encrypted settings, and we suggest this as an area for future research.

We also observed detection and response methods  that were specific to certain problem contexts. In the literature on countering spam, for example, we saw specific interventions that disincentivized its creation by forcing the sender to pay in money or computational work each time a message is sent \cite{EPRINT:JakLinAlg03,cashweb}, taking advantage of the fact that spam is by its nature sent to many recipients simultaneously.
We did not see any other examples of such methods being applied, and we would expect them to be inappropriate for most problem contexts.
For misinformation, two works specifically attempted to detect or limit ``viral'' disinformation \cite{facts,freitas2019can} rather than general misinformation.
We hope to see more research and innovation on this topic both in and out of the encrypted setting.

\begin{figure}
\includegraphics[width=0.95\linewidth]{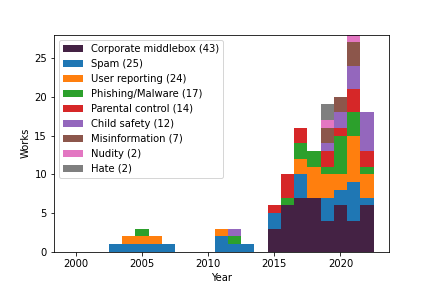}
\caption{E2EE content moderation papers by year and topic. Papers about multiple topics appear in multiple categories.}
\label{fig:results-by-year}
\end{figure}

\end{savenotes}  

\subsection{A need for interdisciplinary work and improved domain expertise}

Figure \ref{fig:results-by-year} shows the moderation goals of all content moderation papers found in our literature search, excluding 155 papers on identifying malware and network anomalies by performing machine learning on encrypted TLS traffic flows.
It shows that most cryptography and security work for exposing rule-violating content was performed in the corporate setting of monitoring TLS traffic for security purposes, with spam mitigation a distant runner-up.
Additionally, aside from a few works with creative response mechanisms we discuss in Section \ref{sec:creative-responses}, very little work unites end-to-end encryption with other potential content moderation responses \cite{goldman2021,gillespie}, such as lowering content visibility or reputation, or (outside the spam setting) disincentivizing its creation in the first place.

We thus call for more interdisciplinary research and
greater attention by the security community to the literature on specific content moderation topics.
We see two barriers to such work.
The first is simply researcher inertia: for years, cryptography had no reason to interact with content moderation topics other than security-adjacent concerns like malware.  Cryptographers and security researchers lack domain expertise in topics like misinformation and child safety that were not traditionally considered security issues, and on the flip side the bulk of social science research on these topics generally either avoids the difficulties associated with encryption, describes them only briefly, or has few answers to share (e.g., \cite{halevy2022preserving,gillespie2020expanding,ganesh2020countering,brown2020regulatory,saha2021short}).
Another obstacle is the difficulty of obtaining data, especially for highly sensitive issues of child safety.

\section{Detection}
\label{sec:detection}

\begin{savenotes}
\begin{table*}
\resizebox{\linewidth}{!}{%
\begin{tabular}{lccccccc}
\textbf{Paradigm} & \textbf{Efficiency} & \textbf{False positive rate} & \textbf{Client privacy?} & \textbf{Server privacy?} & \textbf{Crypto methods}\footnote{See Appendix \ref{sec:crypto} for descriptions of cryptographic objects: Homomorphic encryption (HE), Functional encryption (FE), Private set intersection (PSI), Searchable encryption (SE), Trusted Execution Environments (TEEs).} & \textbf{Threat model} & \textbf{Evasion} \\ \hline
\textbf{Exact matching} & Fast-medium (1ms-2s) & Negligible ($\sim10^{-38}$) & Full & Possible but rare & HE, FE, PSI, or client-side & Usually malicious & Easiest \\
\textbf{Rules/patterns} & Fast (40$\mu$s-400ms)\footnote{The difference between exact and pattern-based matching is likely reflective of the fact that the middlebox setting uses trusted hardware more than other approaches.} & Not determined & Typically partial & Usually & SE, TEEs, or MPC & Usually semi-honest & Moderate-easy \\
\textbf{Perceptual matching} & Similar to Exact & Medium-high ($10^{-8}$-$10^{-3}$)\protect\footnotemark[\getrefnumber{foot:phf-estimate}] & Typically partial & Usually & PSI or client-side & Usually malicious & Moderate-hard \\
\textbf{Machine learning} & Slow (500ms-10s) & High ($10^{-2}$-$10^{-1}$) & Usually & Sometimes &  HE, FE, or client-side & Usually semi-honest & Not determined
\end{tabular}
} 
\caption{Paradigms of automated moderation methods.
All numbers aside from the PHF FPR estimates\protect\footnote{The \emph{worst} estimate of $10^{-3}$ presented here for PHFs' false positive rate is the optimistic \emph{best} performance of all but one PHF provided in Figure 3 of Jain et al.\ \cite{jain2021adversarial}.  The best estimate of $10^{-8}$ is approximately the claimed FPR of Apple's NeuralHash \cite{apple2021ThreatModel}, which is the best FPR of any PHF we know of.\label{foot:phf-estimate}}
are one significant figure of the first and third quartiles from our literature search results.}
\label{tab:summary-of-detection}
\end{table*}

The main body of this section describes the results of our literature search on the detection methods used in different content moderation contexts.
In Section \ref{sec:automated} we discuss automated detection of content, in Section \ref{sec:user-reporting} we discuss methods based on user reporting and on metadata (of \emph{content}, not of \emph{users}, as mentioned in Section \ref{sec:limitations}).
In Section \ref{sec:client-assump} we discuss the fundamental assumptions all these works place on client behavior.

\subsection{Automated detection of content in the literature}
\label{sec:automated}

Most works we examined (all middlebox works and 72\% of non-middlebox works) performed some kind of automated detection of content.   
In Section \ref{sec:problem-context} we discussed how different problem contexts are amenable to different detection methods.
In this section, we examine the technical properties of those different detection methods, including cryptographic methods, accuracy, efficiency, and threat model.
The four main paradigms in content-based detection are:

\begin{enumerate}
\item \textbf{Exact matching}:
typically accomplished with Private Set Intersection, generic Multi-Party Computation, Searchable Encryption, or client-side lists.
\item \textbf{Rule or pattern matches}:
typically accomplished by Searchable Encryption or Trusted Execution Environments (TEEs).
\item \textbf{Perceptual matching}:
typically accomplished by Private Set Intersection or client-side lists.
\item \textbf{Machine learning classification}:
typically accomplished by using Homomorphic or Functional Encryption, or by running the classifier client-side.
\end{enumerate}
We provide a summary of the findings in Table \ref{tab:summary-of-detection}, more details on individual schemes can be found in Tables \ref{tab:individual-nonmiddle} and \ref{tab:individual-middleboxes}.
More details and references for cryptographic methods are in Appendix \ref{app:crypto-methods}.

\paragraph{Exact matching}
\label{sec:exact-matching}
Exact matching refers to systems which attempt to detect content that matches exactly\footnote{Strictly speaking, cryptographic hashing is not ``exact'' matching; the chance of collision is negligible in a designer-chosen security parameter.  However, that chance is typically set around at most $2^{-128} \approx 10^{-38}$.  
In contrast, the lowest false positive rate these authors know of among perceptual hash functions is 3-in-100-million $\approx 10^{-8}$ \cite{apple2021ThreatModel}.
Assuming 4.5 billion non-problematic images shared daily on WhatsApp (a conservative 2017 estimate \cite{jain2021adversarial}) the best perceptual hash function would yield about 135 false positives per day;  
a cryptographic hash would not yield a single false positive for more than the age of the universe in expectation.
Thus, although matching with a cryptographic hash is not perfectly exact, it is exact for all practical purposes and is clearly in a different regime than perceptual hashes.} 
with a particular list of problematic content, typically by comparing the cryptographic hash of the content against the list.
Exact matching
was primarily used for detecting malware-hosting web URLs, malware itself, or performing exact queries within TLS network monitoring.
The TLS middlebox functions generally provided partial client privacy \cite{jia2022encrypted,lin2016privacy},
but all other systems provided full client privacy.

The server-private works mainly achieved their goals via homomorphic or functional encryption, or general cryptography protocols (see Appendix \ref{sec:crypto} for more information on cryptographic tools).
Those that did not provide server privacy mainly faced a technical challenge of compressing the information to be stored on the client device as much as possible.
The works that used exact matching typically had extremely low false positive rates, typically 0 or negligible in the cryptographic sense.
Those works that did have higher false positive rates typically made use of a bloom filter \cite{bloom} with a tunable parameter for false positives; these also had full client privacy, avoiding privacy issues from a false positive.

We did not include certificate transparency in our SoK since it is a moderation of identities rather than content.  However, certificate transparency often performs fully client-private exact matching with negligible or zero false positives, often by pushing as much as possible to the client \cite{CCS:SSWKQ21,NDSS:SmiDicSea20,SP:LCLMMW17,CCS:SchLevSpr14,peeters2013privacy,narasimha2007privacy,PETS:SolTsu06}.

\end{savenotes} 

\paragraph{Rules and patterns}

A second approach to content moderation in E2EE
is to set a rule, predicate, pattern, or other simple search query, and to selectively reveal or block messages or traffic whose plaintext matched the search.

This detection approach was overwhelmingly used in TLS middleboxes implementing deep packet inspection via searchable encryption or trusted hardware; see Appendix \ref{sec:big-tables} for details.
Less frequently, this approach was also used for other use cases, including text filtering \cite{PKC:NSSWXZ21}, and detection of spam \cite{bian2019towards,zhang2015pif}, malware and phishing \cite{whatsappsuspiciouslinks,jammalamadaka2005pvault}, and hate speech \cite{ramezanian2019privacy,zhang2013safe}.

The very core of policy-based moderation, of course, presupposes that the service provider has some relatively concise description of the policy the clients should follow.
In some cases, this could become a short list for matching. A service provider could, for example, list keywords or specific content that the matching protocol would detect and disallow.
However, a rule or pattern can often achieve more efficient results than a large list.  
Lists and keywords, when searched directly, are also easily avoidable without more sophisticated learning \cite{SP:ZGBWCFB21} and many of the papers we will discuss shortly about ML-based detection describe the inadequacy of policy-based detection for many tasks.

Unfortunately, the technologies most frequently used to implement this paradigm---searchable encryption and trusted execution environments---have some drawbacks that would require care to integrate with full client privacy.  
Searchable encryption frequently exhibits leakage that is unacceptable for E2EE (e.g., \cite{kamara2022sok,USENIX:OyaKer21,NDSS:IslKuzKan12}),
and only some searchable encryption schemes are compatible with forward secrecy  \cite{bost2016ovarphiovarsigma,zhang2019towards}.
Trusted hardware also has many known side channel attacks, many of which are known to fully exfiltrate encryption keys \cite{nilsson2020survey,chen2019exploitable,van2018foreshadow,chen2018sgxpectre,lindell2018security,gotzfried2017cache}.
In the enterprise setting, where it is common for middleboxes to fully decrypt TLS traffic and read plaintext packets \cite{de2020survey}, this privacy leakage presents only moderate concern. But these issues are a major problem in E2EE settings where privacy is the norm.

\paragraph{Perceptual matching}
\label{sec:phf}

A challenging aspect of content moderation is that senders of problematic content will often try to evade the detection mechanism.
Users bypass word filters by misspelling words \cite{grondahl2018all,papegnies2017detection},
and for images and video users use common methods to evade matching: altering a small number of pixels, changing the size, rotating the image, or changing the aspect ratio \cite{farid2021overview,dalins2019pdq}.  These approaches will completely alter the cryptographic hash of an image or message, but they may not interfere much with human perception of the content.

The response to this evasion is the \emph{perceptual hash function} (PHF) \cite{venkatesan2000robust}.
PHFs are locality-sensitive hashes \cite{DBLP:conf/vldb/GionisIM99} that return the same or similar hash values even if the input has been put through a class of perturbations.
A variety of PHFs are used in industry for content moderation, primarily for images and video. These include Microsoft's PhotoDNA \cite{photodna}, Facebook's PDQ for images and TMK+PDQF \cite{facebookPdq} for video, and Apple's proposed PHF for CSAM called NeuralHash \cite{apple2021csam}.
The academic literature contains more PHFs \cite{ du2020perceptual,biswas2021state,farid2021overview,phash,norouzi2012fast,neelima2016perceptual}.

PHFs do \emph{not} aim to achieve the same level of collision resistance as cryptographic hash functions (CHFs),
instead they aim to provide collision resistance for images that are not perceptually similar \cite{weng2011secure}.
Unsurprisingly given their goals, PHFs have higher false positive rates than CHFs even for unrelated images \cite{jain2021adversarial,steinebach2012forbild,DBLP:conf/ccs/HaoLJ021}, between 1-in-1000 \cite{jain2021adversarial} to 1-in-10-million \cite{DBLP:conf/cvpr/SinghF19} to 3-in-100-million \cite{apple2021ThreatModel}.
See Appendix \ref{sec:phf-attacks} for a summary of recent attacks on PHFs, and for benchmarks of common PHFs, see \cite{thornPhfBenchmark}.

We saw very few examples of PHF-based content moderation in E2EE in our literature search.  PHFs appear only in Reis et al.'s 2020 work for misinformation in WhatsApp that provides full client privacy \cite{reis2020detecting} and the two 2021 partially client private proposals for matching CSAM \cite{bhowmick2021apple,USENIX:KulMay21}.
A few more papers we examined use locality-sensitive hashes for identifying spam similar to previously-seen spam \cite{yao2022privacy,tarafdar2021spam,damiani2004p2p}, however aside from these PHFs are rare in the literature we examined.
We hope to see both improved PHFs and improved scrutiny of PHFs in the future.

\paragraph{ML Classification}
Our final category of automated content-based detection is machine learning (ML) classification.
Over our entire search, 24\% of papers performed privacy-preserving ML to do a content moderation task, spread across the moderation goals of improving security,
spam,
and other topics (see Tables \ref{tab:individual-nonmiddle} and \ref{tab:individual-middlebox}).
The year 2021 saw newfound activity for using ML to detect CSAM in various forms:
Four of the five contest entries to the recent U.K. Safety Tech Challenge \cite{ukSafetyTechWinners} utilized a client-side ML model to detect CSAM  \cite{galaxkey,dragonflai,t3k,safetonet}, including self-generated CSAM \cite{safetonet}, and Apple's Communication Safety uses ML detecting nude pictures in children's chat messages \cite{appleCommunicationSafety} with full client privacy.

Our accuracy findings (see Table \ref{tab:summary-of-detection}) provide evidence toward the notion that ML typically has a higher false positive rate than matching via perceptual hash functions \cite{levyRobinson},
however further work is needed to see if these results remain true for the best classifiers.

Of the machine learning based detectors in our search, 54\% used some form of cryptographic protocol to aid private computation of the machine learning, and
39\% were implemented client-side.

All but three ML-based designs maintained full client privacy; those three were in especially controlled settings of enterprises or parental control\cite{ramezanian2021parental,ramezanian2019privacy,alabdulatif2017privacy}. (Three UK Safety Tech Challenge entries were agnostic as to the client privacy setting \cite{galaxkey,dragonflai,t3k}.)

\subsection{Approaches that do not rely on automated content detection}
\label{sec:user-reporting}

Just under half of the non-middlebox works (46\%) incorporated user reporting or analysis of content-agnostic metadata.

\paragraph{User reporting}
User reporting was a primary detection mechanism of 33\% of non-middlebox works.
It was used most frequently in the context of mis/disinformation and general reporting of abusive messages.
For mis/disinformation, one key area was tiplines or monitors for fake news on WhatsApp 
\cite{martins2021fact,kazemi2021tiplines,meedan2020one,bagade2020kauwa,melo2019whatsapp}.
Some spam and malware works also relied on user detection to identify malicious messages, then blocked future copies automatically.

In Section \ref{sec:message-franking} we will discuss \emph{message franking}, an important component of E2EE content moderation which adds cryptographic verification to the process of user reporting, ensuring that malicious receivers cannot frame senders for content they did not send, and honest receivers can prove a sender really did send a particular message.  These works are displayed with the goal of User Reporting (UR) in Table \ref{tab:individual-nonmiddlebox}, but we defer in-depth discussion of these works to Section \ref{sec:message-franking} since they implement an accountability property on the existing detection mechanism of user reporting.

\paragraph{Metadata-based measurement}
14\% of non-middlebox works incorporated some analysis of metadata about content,
as opposed to (or in addition to) analysis of the content itself.
These fall into two main categories:
One group identifies identifying spam or phishing alerts;
these works attempted to discourage or block spam based on volume or the existence of particular links.
The other group of metadata-based works were those that attempted to detect encrypted files containing CSAM by performing machine learning on filename metadata \cite{pereira2020metadata,al2020file,panchenko2012detection}.
Some works also incorporated metadata-based analysis alongside content-based analysis.

We encourage research in new methods using non-content sources, and research measuring the efficacy, accuracy, and other properties of user reporting and metadata-based measurement compared to each of the automated content-based detection paradigms.

\subsection{Assumptions on client behavior}
\label{sssec:threat:uibound}
\label{remark:client-threat}
\label{sec:client-assump}

Any detection mechanism for catching clients who are trying to evade detection requires at least some assumptions beyond the cryptographic threat model.
At one extreme, it is impossible to thwart a sufficiently motivated and sophisticated colluding sender and receiver from using the channel to send problematic content: the sender and receiver could run their own key exchange on top of the existing channel and build their own layer of encryption on top of the existing one.  Any detection scheme capable of detecting the encrypted content would be able to break encryption generally.
However, adding this additional layer of encryption requires a good amount of technical sophistication and cooperation on the part of both the sending \emph{and receiving} client; in situations where both the sender and receiver are colluding, we conjecture that malicious senders would prefer to send perceptually-recognizable problematic content (although we know of no research into this question).

Much of the middlebox literature explicitly assumes the existence of one honest client as a core requirement, to avoid this exact problem and assume away any malicious out-of-band pre- or post-processing \cite{sherry2015blindbox}.
The appropriate choice of client threat model is highly dependent on the problem context.

The remaining automated detection options become a ``cat-and-mouse'' adversarial game:
the platform tries to cast a wide enough net to catch users exchanging problematic content without catching unacceptably many false positives, and malicious users try to modify the content they send just enough so that it evades detection but is still recognizable.
The existence of perceptual hash functions
is a concession to this cat-and-mouse game: 
PHFs have high false positive rates compared to exact matching, but are harder to evade.

If PHFs become a key feature of content moderation under E2EE, then their improvement and analysis will lead directly to improved accuracy, simultaneously reducing the privacy loss, increasing the difficulty of evasion, and increasing the difficulty of maliciously induced false positives (i.e., where a user sends an image that appears innocuous but that has the same hash as harmful media).
Recent work analyzes the security properties of perceptual hash functions \cite{farid2021overview,prokos2021squint,jain2021adversarial,dalins2019pdq,drmic2017evaluating} (see Appendix \ref{sec:phf-attacks} for more) and we encourage more research on this front.

\section{Response}
\label{sec:response}

In this section we briefly describe the results of our search on client privacy (see Appendix \ref{sec:client-privacy} for more details) and then
we go into the response mechanisms we saw in the literature that are unique to E2EE and irrelevant for standard content moderation.
We see fruitful areas of future research there which go beyond the proposals to improve detection in a more obvious way.

\subsection{Client privacy in moderation response}
Of the non-middlebox works for which client privacy was relevant and we could identify the a client privacy setting, 88\% offered \emph{full} client privacy.
At the same time, within the TLS middlebox literature, 98\% of middlebox designs offered \emph{partial} client privacy.

This difference is stark.
We see at least two factors that explain this gap.  
First, as we observed in our discussion of rule-based detection in Section \ref{sec:detection}, it is common for non-privacy-preserving middleboxes to break the TLS connection entirely.  This makes partial client privacy a step up in privacy rather than a step down, as it would be for most content moderation under E2EE.
Second, in the corporate settings where middleboxes are frequently used, there is often an expectation that all activities are monitored that is absent in typical E2EE deployments.  A 2017 survey by O'Neill et al.\ \cite{oneill2017tls} shows much stronger public support for general TLS proxies in the corporate setting --- and to a lesser extent, schools --- than any other context.
Understanding these factors, as well as any other considerations that help choose full or partial client privacy for different situations, is a useful area for future research.
See Appendix \ref{sec:client-privacy} for more details on the client privacy difficulties for different detection paradigms.

\subsection{Responses unique to end-to-end encryption}
\label{sec:creative-responses}
In the non-encrypted setting, already a wide variety of content moderation responses exist, including banning, suspending, lowering visibility, fining or withholding money, and so on \cite{goldman2021}.

Most papers in our literature search handled content moderation responses in the same way the issue would be handled in a standard content moderation setting: either informing the server or a moderator of the detection (allowing whatever actions the server deems appropriate),
warning the client of the detection (as in malware), or invisibly sending the content to another folder until the client re-identifies it (as with spam).

However, some literature utilized specific information about the E2EE setting that enabled new responses that are not applicable in the standard setting.
These mainly revealed new information about a previously-encrypted message, once it has been detected by user reporting or automated methods.

Peale et al.\ \cite{CCS:PeaEskBon21}, and later Issa et al.\ \cite{USENIX:IssaAlHVar22}, implemented \emph{source tracking}.  Source tracking effectively encodes the original source of a sender into a message: if a message is originally sent from A to B, B forwards it to C, and C reports it, the service provider will learn that A was the original sender.
Peale et al.\ also create an extra confidentiality property of the forwarding path: if a client receives the same message from two different sources, the ``tree-unlinkability'' property ensures that the client will not know whether the message was received via the same forwarding path both times.

Tyagi et al.\ \cite{CCS:TyaMieRis19} implemented \emph{traceback} for E2EE messaging:
after a detection,
the service provider gains the ability to ``trace'' the forwarding path the message took to get to the receiver in one of two ways.
Suppose A sends a message to B and C.  B forwards the message to D, and separately, C forwards the message to E.  E later reports the message.
Under \emph{path forwarding}, the service provider learns the message path $A \rightarrow C \rightarrow E$, and could take action on that path (e.g., in the setting of misinformation, could warn the users after the fact that the information was suspicious).
Under \emph{tree forwarding}, the service provider learns the paths $A \rightarrow C \rightarrow E$, and also the path $A \rightarrow B$, though not the fact that B forwarded the message to D.  

In a different take on user reporting, Liu et al.\ \cite{facts} described an approach for revealing messages once they reached a specific \emph{threshold} of reports globally across all users, by constructing a ``collaborative counting bloom filter.''  Their goal was to reveal misinformation that was especially ``viral'' and thus by definition reached a large number of users; after this the server would be free to take action on those specific images (e.g. by sending it to client devices to perform matching to append a warning if it is seen again by future clients).
This is reminiscent of earlier schemes that perform similar goals for spam \cite{kong2005scalable,damiani2004p2p}.

We encourage more researchers to think outside the box of ``binary detection,'' which captures the majority of literature, and to continue exploring content moderation responses that could respond to societal harms while respecting full client privacy.

\section{Transparency and accountability in content moderation}
\label{sec:transparency}
\label{sec:server-accountability}

One of the strongest criticisms of the 2021 Apple CSAM detection tool was the risk that the system would inevitably bow to pressure to expand the use cases of the surveillance system to other purposes \cite{biop,cdtLetter,aclu,eff2021,vergeApple,rescorla2021more,wapo,nyt,corfield2021apple,newamerica,cato}.
As others have pointed out \cite{cdt}, once the plaintext can be processed for one purpose (detecting CSAM), additional exceptions could be carved out for other purposes (e.g., detecting threats to national security, terrorist content, hate speech, misinformation, or more), or the system could be exploited by external malicious parties to undermine privacy in other ways.
There is also likely to be significant international pressure to censor or track specific political memes \cite{cdtLetter,appleNYT,appleNavalny}.

The tension between free speech and content moderation was already an issue in non-encrypted content moderation \cite{gillespie,grimmelmann2015virtues,klonick2020content,dias2020content,langvardt2017regulating,gorwa2020algorithmic,douek2022siren}. 
However, the stakes are higher in content moderation that bypasses encryption, because encryption is one of the few methods by which over-broad surveillance can be avoided.
Thus, under encryption, maintaining transparency, oversight, and verification of content moderation is paramount.

Numerous transparency mechanisms have been proposed and enacted in the non-encrypted setting, including
community guidelines and terms of service,
aggregate transparency reports \cite{googleTransparencyReport,facebookTransparencyReport},
legal or contractual boundaries \cite{langvardt2017regulating,douek2022siren},
oversight boards \cite{fboversight},
third-party audits (e.g., \cite{murphy2020facebook,bhowmick2021apple}),
and a variety of other governance approaches \cite{de2020democratising,douek2022siren}.
These approaches should be applied in E2EE content moderation as well,
though they also have recognizable limits even without encryption \cite{gorwa2020algorithmic}.

We propose that cryptographic means of enforcing transparency be utilized not only in the E2EE setting but also the unencrypted setting.
Section \ref{sec:transparency-literature}  describes transparency mechanisms we saw in our literature search,
and in Section \ref{sec:transparency-research} we propose future research.

\subsection{Transparency methods in the literature}
\label{sec:transparency-literature}

\paragraph{Verifying the server}

In the Transparency column of Tables \ref{tab:individual-nonmiddlebox} and \ref{tab:individual-middlebox}, we identify any transparency properties by which the client can verify the system's correct behavior.

We identified 82 works where the client must rely on some promise or information held by the server (e.g., a secret dataset or model, or an honesty assumption).
Of these, 49\% made at least some mention of the need for transparency.
Of the works that mentioned some form of transparency goal, 30\% provide no concrete guidance on how to achieve it.
An additional 20\% make an explicit assumption of honesty on some party, typically a ``rule generator'' party with no other input.   A small number of works mention third-party audits (5\%), or getting consent from clients (5\%).

Works that use trusted execution environments often mention attestation as a means of establishing transparency (77\%), however in all but one case \cite{schlogl2020ennclave}, the attestation would only be verifiable to the server, not the client.

The remaining four works had some concrete actionable transparency proposal:
Section 4.4 of the Pretzel spam detector for E2EE by Gupta et al.\ \cite{gupta2017pretzel} is dedicated to transparency issues.  In addition to preventing all but one bit of leakage against a malicious server, Gupta et al.\ also discuss a particular client action that would allow the client to ``opt out with plausible deniability'' by garbling the incorrect function without the server's knowledge.
Second, Apple's PSI proposal for detecting CSAM \cite{bhowmick2021apple} suggested cryptographic methods for verifying the server's set that could be implemented by a third-party auditor in a secure environment to ensure that the content moderation system only relied on CSAM hashes from child safety groups \cite[p. 13]{bhowmick2021apple}.
The eNNclave work \cite{schlogl2020ennclave} made use of the trusted hardware attestation functionality as well, but the client checked the code rather than the service provider.
Fourth, SAFE \cite{zhang2013safe} describes a protocol in which clients share hashes of items they wish to filter (e.g. hate or spam).  The protocol used a Merkle tree \cite{merkle1987digital} to authenticate the filters.

We also know of two works that build novel cryptographic transparency mechanisms which became available in preprint after the conclusion of our literature search.
First, Bartusek et al.\ \cite{bartusek2022end} offer
the ability to enforce a predicate on the class $C$ that forms the partial client privacy exception.
They describe constructions of ``set-pre-constrained'' group signatures and encryption in which all parties can verify, for example, that a list $C$ used for matching is capped at size at most $n$, or other predicates about $C$.
Second, Scheffler et al.\ \cite{SP:SchKulMay23} perform policy analysis of partially client-private systems using exact or perceptual matching for CSAM, and suggest three protocols to improve their transparency: (1) use threshold signatures among child safety groups providing hash sets, (2) allow the server to prove that particular elements are \emph{not} in the hash set, and (3) ensure that users with matching content (true or false positive) eventually learn that their content was revealed, after a delay.

In general, we see cryptographic transparency and auditability methods as a useful area of future research: many of the works mentioned the importance of transparency, but few had concrete methods for allowing the client to verify the server's behavior.
Since this topic is also at the heart of the debate over content moderation in E2EE generally,
we believe it is a worthy research agenda within both technical and non-technical approaches.

\paragraph{Fully client-side content moderation}

Some works avoid the problem of verifying the server's behavior by avoiding the server's direct involvement at all parts of the content moderation pipeline, and thus control of the scheme is essentially always held by the client.  In these settings, false positives or negatives may cause other problems such as being unable to view important messages, but no privacy was lost.
Many of these works also state or imply that they are meant to be ``soft'' moderation methods: the client is able to bypass the categorization if they wish (e.g. they can view the ``spam folder'' and remove items from it).
In these designs, code inspection and continued use should demonstrate the correct functionality of the system; there is no remote server that needs periodic inspection or auditing.
A privacy improvement can also be achieved even for partial client privacy, by reducing the amount of information that is sent to the server \cite{jin2022peekaboo} or by performing detection only after a threshold of problematic content was detected \cite{bhowmick2021apple,facts}.

A future research line fusing the literature on E2EE content moderation, verifiable programming, privacy, and systems security would help develop the transparency and security properties needed for content moderation running fully on client devices.

\paragraph{Authenticating client reports}
\label{sec:message-franking}

In addition to verifying the server's behavior, one strong area of cryptography research in the literature is in the realm of verifying client behavior during user reporting: users are cryptographically prevented from forging a user report that would frame an innocent sender.
Technically speaking,
\emph{message franking} is a three-party protocol between a sender, receiver and moderator that adds two additional accountability properties to user reporting, at the cost of some deniability.  In a message franking scheme, senders always attach a ``signature'' to every message sent, in such a way that if a sender sends problematic content to a receiver, that receiver can report the message to a moderator, who will check the signature to ensure the sender truly sent it.  This cryptographically ensures that receivers cannot report to the moderator messages that are ``forged'' to appear as if they were from the sender; the moderator cannot be convinced any party sent a message they did not send.  These schemes have two key accountability properties \cite{C:GruLuRis17,C:TGLMR19}:
\begin{enumerate}
\item \emph{Sender binding}: If a sender sends a message that can evade the moderator's verification, the receiver will refuse to validate the message at all, treating it as malformed.
\item \emph{Receiver binding}: The receiver is unable to forge the sender's signature to the moderator.
\end{enumerate}
These two properties together also ensure that no one can impersonate a sender to the receiver \cite{C:TGLMR19}.

Message franking is a key component of user reporting in E2EE secure messaging. 
User reporting is an attractive option in E2EE because one of the ``ends'' of the communication must take positive action before any new information is revealed to the server.

These accountability properties come at a subtle cost to the typical deniability property of E2EE (see Section \ref{sec:e2ee}).  Under the proposed designs of message franking 
the E2EE deniability property will no longer hold against the moderator, although it will still hold against third parties who do not know the moderator key.  See the full version of \cite{C:TGLMR19} for variants of message franking with different deniability properties.
All the message franking schemes we reviewed in Section \ref{sec:litsearch} achieve some variant of these transparency properties, 
including some schemes that are compatible with sender-private networks 
and some that achieve the forward secrecy property of E2EE in addition to accountability.

\subsection{Suggested future research in transparency}
\label{sec:transparency-research}
We saw two key areas for future research in transparency that were not well-examined in the current literature.

\paragraph{Privacy-preserving protocols for aggregate statistics}
Prior work has stressed the difficulties that E2EE will pose on measuring the accuracy and effectiveness of a content moderation system (e.g. \cite{halevy2022preserving,kardefelt2020encryption}).
Service providers are understandably nervous about losing the ability to measure the aggregate performance of their systems under E2EE; a reason to avoid E2EE in the first place.

We propose that \emph{privacy-preserving aggregate telemetry and measurement} of the content moderation system is warranted and helpful for both clients and service providers alike.  These systems could be based on methods for secure E2EE telemetry \cite{facebookDIT,corrigan2017prio,PETS:MBGKS08}, federated learning \cite{hard2018federated}, accountability in other settings \cite{frankle2018practical}, or methods for privately measuring aggregate information in Tor \cite{PoPETS:CDKY20,cotacallapa2020measuring,CCS:WJSYG19,CCS:FMJS17,NDSS:MelDanDeC16,CCS:JanJoh16}.

Giving service providers access to telemetry and aggregate statistics will also enable measurements and improvements of other aspects of the system like algorithmic fairness \cite{DBLP:conf/innovations/KleinbergMR17,DBLP:journals/bigdata/Chouldechova17}, allowing continuation of techniques used in the non-encrypted setting \cite{googleFairness,facebookFairness}.
The capability of gaining aggregate statistics about
a detector is also likely to make the implementation of E2EE and other privacy-preserving systems more palatable for services currently on the fence about providing E2EE.

We believe the development of these systems would serve several purposes.
Not only would they allow service providers to monitor and improve their content moderation systems in an aggregate way, they would also help clients verify certain claims about the content moderation system, such as its false positive rate on a global scale.  The ability for the public to audit or verify these claims is a key principle in this and other areas of cryptography policy \cite{carnegiePrinciples,biop}.

\paragraph{Enabling and enforcing notice, appeal, and redress in E2EE systems}
The Carnegie principles and Abelson et al.\ \cite{carnegiePrinciples,biop} detail the importance of the principle of ``Accountability: When a phone is accessed, the action is auditable to enable proper oversight, and is eventually made transparent to the user (even if in a delayed fashion due to the need for law enforcement secrecy)'' \cite{carnegiePrinciples}.
In contrast to the aggregate transparency mechanisms suggested above, we also note that individual notice to users who have had their content moderated is an important aspect of moderation.  Providing explanations for content removal above and beyond the fact of removal itself has also been found to improve user behavior in the future \cite{jhaver2019does}.
Appeal is also critical for any content moderation system,
and reporting moderation decisions to the user is an important prerequisite for any appeal and redress mechanisms.

These ideas receive very little attention in the technical literature on E2EE content moderation.
Partially client-private systems which report detections to the server rather than client sometimes have no mechanism  in place---cryptographic or otherwise--- for ensuring the client receives notice for her moderated content, let alone the ability to appeal it.

For automated systems in E2EE, appeal also poses a technical challenge: if a sender sends a benign message that is falsely flagged as problematic content, and the content is reported to a human moderator who determines the content is not problematic, then what technical means should be taken to ensure the client is able to send their message as soon as possible?  
Naive solutions like allowlists leave many privacy and efficiency issues unanswered, and so we encourage future work in this area as well.

\section{Conclusion}
\label{sec:conclusion}

Grimmelmann summarized the difficulty of content moderation by saying that ``responsible content moderation is necessary and $\ldots$ responsible content moderation is impossibly hard'' \cite{grimmelmann2017platform}.
The same is doubly true on both counts under end-to-end encryption:
encryption allows people to hide bad behavior from reasonable moderation, but also remains one of the only bastions against unreasonable government and corporate surveillance.

Although there will likely never be a perfect content moderation system, let alone one operating under E2EE, the current systems leave much to be desired and have tractable problems that can be addressed with future research over the coming decade.
We hope our work provides a foundation on which to do further research that will enable forward progress towards this demanding goal.

\begin{acks}
Many thanks to Anunay Kulshrestha for assisting in manual extraction and confirmation of information from the works chosen for analysis in the literature search.  Additional thanks to Andrew Fishberg for assistance in the process of downloading and running automated extraction of forward and backward citations in the initial phase of the literature search.
We also thank the anonymous reviewers for their high-quality feedback and suggestions for improvements to this work.
This work was supported by the Center for Information Technology Policy at Princeton University, and also by a cryptography research grant from Ripple.
\end{acks}

\appendix
\section{Literature search details}
\label{sec:methods}
\label{app:methods}

This section contains additional details on our literature search beyond those in Section \ref{sec:lit-search}.

The initial sample for our literature search was created by running the following queries in August 2022 in computer science and cryptography-related academic venues: content moderation, CSAM, end to end, malware, misinformation, porn, pornography, and spam.
The academic venues were 
ACM CCS,\footnote{\url{https://dl.acm.org/conference/ccs/proceedings}}
CRYPTO,\footnote{searched via DBLP \url{https://dblp.org/} and filtered by venue\label{foot:dblp}}
NDSS,\footnotemark[\getrefnumber{foot:dblp}]
PETS,\footnotemark[\getrefnumber{foot:dblp}]
IEEE S\&P,\footnote{\url{https://ieeexplore.ieee.org/Xplore/home.jsp} filtered by venue}
Usenix Security,\footnote{\url{https://www.usenix.org/publications/proceedings}}
arXiv CS,\footnote{\url{https://arxiv.org/search/cs}}
and
IACR ePrint,\footnote{https://eprint.iacr.org/search}.
We additionally examined the top 200 results from Google Scholar for ``encrypted content moderation'' and ``end to end encrypted content moderation.''
For the computer science venues, we iteratively refined query terms per venue if the search returned too many results to manually search and inspection revealed that the results were irrelevant (e.g. a search for ``content moderation'' would surface matches for ``content'').
The final queries to form the initial set, with their result counts, were as follows:
\begin{itemize}
\item ACM CCS: ``content moderation'' (4), CSAM (1), ``end to end encryption'' (61), misinformation (26), porn (16), pornography (42), ``privacy-preserving'' AND ``malware-detection'' (10), ``private'' AND ``malware detection'' (51), spam AND encryption
\item CRYPTO: content moderation (1), CSAM (0), end to end (2), malware (0), misinformation (0), porn (0), pornography (0), spam (1)
\item NDSS: content moderation (0), CSAM (0), end to end (3), malware (30), misinformation (0), porn (0), pornography (0), spam (7)
\item PETS: content moderation (0), end to end (1), malware (1), misinformation (0), porn (0), pornography (0), spam (0)
\item IEEE S\&P: content moderation (1), CSAM (0), end to end encryption (17), misinformation (0), porn (0), pornography (0), child pornography (0), privacy malware (84), spam (9)
\item Usenix Security: ``content moderation'' (1), CSAM (0), ``end to end encryption'' (8), ``malware detection'' (9), misinformation (0), porn (0), pornography (1), spam (27)
\item arXiv CS: ``content moderation'' (74), CSAM (10), ``end to end encryption'' (72), encryption misinformation (5), porn (14), pornography (31), privacy-preserving malware detection (9), private malware detection (20), encryption spam (11)
\item IACR ePrint: content moderation (428), CSAM (0), ``end to end encryption'' (31), ``end to end'' (173), malware (32), misinformation (3), porn (0), pornography (0), child pornography (7), spam (28)
\item Google Scholar: encrypted content moderation (200 examined), end to end encrypted content moderation (200 examined)
\item UK Safety Tech Challenge \cite{ukSafetyTechWinners}: 
Details of the five winners  
obtained from the End of Programme Supplier Showcase event \cite{eopUKSTC}.
\item The following documentation for E2EE services: the  Apple child safety page on August 5, 2021,\footnote{\url{https://web.archive.org/web/20210805200549/https://www.apple.com/child-safety/}}
Google Messages support,\footnote{\url{https://support.google.com/messages/}}
Signal support,\footnote{\url{https://support.signal.org/hc/en-us}}
WhatsApp Help Center\footnote{\url{https://faq.whatsapp.com/}}
\end{itemize}

As mentioned in Section \ref{sec:lit-search}, 
we manually examined papers to identify relevant works.  
To be considered relevant, works must:
\begin{enumerate}
\item Include at least one subsection on a content moderation method, system, implementation, or construction, and
\item Provide either partial client privacy with respect to a class corresponding to the content moderation problem, or full client privacy (Definitions \ref{def:full-client-privacy} and \ref{def:partial-client-privacy}).  For systems that can be run client-side, they must mention that they are intended to be used in an encrypted or private setting.
\end{enumerate}

Once the initial set of relevant papers was identified, further works were identified from those works via snowball sampling.
Citations were scraped via the publisher's API when available, 
then via Semantic Scholar or arXiv if located there,
or extracted from the PDF paper itself if no other version was available.
Forward references were found via Google Scholar's ``cited by'' feature.

\subsection{Excluded works}
The researchers excluded papers on topics adjacent but not identical to content moderation under E2EE.  The researchers excluded many papers on the topic of detection/blocking of misbehaving users in anonymous networks, certificate transparency and revocation checking, moderation of the blockchain, implementations of Digital Rights Management (DRM) or watermarking schemes, key escrow or access control schemes, traitor tracing, or measurement of E2EE systems without moderation.

In the case that multiple versions of the work were identified, we only included one version.  We included journal papers over conference papers, and conference papers over preprints or manuscripts.
We also excluded any work where we could not get access either publicly, via institutional login, or through a loan to the Princeton University Library, and works that were not in English.

Finally, as discussed in Section \ref{sec:e2ee}, we excluded works that did not meet Definition \ref{def:partial-client-privacy} for results not in the problematic class $C$, in particular this means we excluded works that sent raw cryptographic or bloom filter hashes of all content to the server.
\section{Moderation of TLS traffic at middleboxes}
\label{sec:detailed-tables}
\label{sec:big-tables}

Tables \ref{tab:individual-nonmiddle} and \ref{tab:individual-middleboxes} show our detailed examination of each work we examined for our full literature search.
Table \ref{tab:individual-nonmiddle} in Section \ref{sec:current} shows all non-TLS middlebox works.  Here, Table \ref{tab:individual-middleboxes} shows the works on middleboxes.

\begin{savenotes} 
\begin{table*}
\resizebox{\linewidth}{!}{%
\begin{tabular}{ll|ccc|ccc|lll|cc|cc|cl}
\multicolumn{2}{c|}{\textbf{\ul{Problem context}}} & \multicolumn{11}{c|}{\textbf{\ul{Detection}}} & \multicolumn{2}{c|}{\textbf{\ul{Response}}} & \multicolumn{2}{c}{\textbf{\ul{Transparency}}}\\
\textbf{Work} & \textbf{Goal} & \rot{\textbf{Exact matching}} & \rot{\textbf{Rules/patterns}} & \rot{\textbf{Machine learning}} & \rot{\textbf{General Crypto/MPC}} & \rot{\textbf{Trusted Hardware}} & \rot{\textbf{Searchable Encryption}} & \rot{\textbf{Size of largest ruleset}} & \rot{\textbf{Latency}} & \rot{\textbf{Communication}} & \rot{\textbf{Security Against Server}} & \rot{\textbf{Security Against Client}} & \rot{\textbf{Server privacy}} & \rot{\textbf{Client privacy}} & \rot{\textbf{Transparency}} & \textbf{Details}\\
\hline
Yao et al.\ \cite{yao2022privacy} & General middlebox & \fullcirc &   &   &   & \fullcirc &   & 5M & 0.5ms & $\times$ & Mal. & $\times$ & \fullcirc & \halfcirc & \emptycirc &   \\ 
Grubbs et al.\ \cite{USENIX:GAZBW22} & General middlebox &   & \fullcirc &   & \fullcirc &   &   & 2M & 3.1s &        & S.H. & Mal. & \emptycirc &   & \quartercirc & Mention \\ 
Ren et al.\ \cite{ren2021enabling} & General middlebox &   & \fullcirc &   & \fullcirc &   & \fullcirc & 3k & 287s & 0.4 MB & S.H.(NC) & $\times$ & \fullcirc & \halfcirc & \quartercirc & Mention \\ 
Lai et al.\ \cite{lai2021practical} & General middlebox &   & \fullcirc &   & \fullcirc &   &   & 24k & 0.85ms & 5x & S.H. &   & \fullcirc & \halfcirc & \emptycirc &   \\ 
Fan et al.\ \cite{fan2020group} & General middlebox &   & \fullcirc &   & \fullcirc &   & \fullcirc & $\times$ &  Asymp. & $\times$ & $\times$ & Mal. & \fullcirc & \halfcirc & \emptycirc &   \\ 
Desmoulins et al.\ \cite{desmoulins2018pattern} & General middlebox &   & \fullcirc &   & \fullcirc &   & \fullcirc & 6k & 2.3ms & 64x & Mal. & S.H. & \emptycirc & \halfcirc &   &   \\ 
EndBox\ \cite{goltzsche2018endbox} & General middlebox &   & \fullcirc &   &   & \fullcirc &   & 377 & 1.67ms & 16\% & S.H. & Mal. & \fullcirc & \halfcirc & \halfcirc & Attestation \\ 
SPABox\ \cite{fan2017spabox} & General middlebox &   & \fullcirc &   & \fullcirc &   &   & 3k & 0.5ms &        & S.H. &   & \fullcirc & \halfcirc & \quartercirc & Mention \\ 
Zhou and Benson\ \cite{zhou2015towards} & General middlebox & \fullcirc &   &   &   &   &   & $\times$ & 10s & $\times$ & S.H. &   &   & $\star$\footnote{Reveal public but not private content} & \emptycirc &   \\ 
Li et al.\ \cite{li2022towards} & Intrusion detection &   & \fullcirc &   & \fullcirc &   & \fullcirc & 1.6k & 6.5ms & $\times$ & S.H. & Mal. & \fullcirc & \halfcirc & \quartercirc & Mention \\ 
Chen et al.\ \cite{chen2022privacy} & Intrusion detection &   & \fullcirc &   & \fullcirc &   & \fullcirc & 3k & 103ms & 407 KB & S.H. & Mal. & \fullcirc & \halfcirc & \quartercirc & Honesty assumption \\ 
Jia and Zhang\ \cite{jia2022encrypted} & Intrusion detection & \fullcirc & \fullcirc &   & \fullcirc &   & \fullcirc & 3k & 267s & 82 MB & $\times$ & Mal. & \fullcirc & \halfcirc & \quartercirc & Honesty assumption \\ 
Chuchotage\ \cite{nikbakht2022chuchotage} & Intrusion detection &   & \fullcirc &   &   & \fullcirc &   & $\times$ & 0.07s & $\times$ & TEE & Mal. & \fullcirc & \halfcirc & \halfcirc & Attestation \\ 
Canard and Li\ \cite{canard2021towards} & Intrusion detection &   & \fullcirc &   & \fullcirc &   & \fullcirc & 3k & 1.5us & $\times$ & S.H. & Mal. & \fullcirc & \halfcirc & \emptycirc &   \\ 
Guo et al.\ \cite{guo2020privacy} & Intrusion detection &   & \fullcirc &   & \fullcirc &   & \fullcirc & 1.6k & 5ms & 400 B & S.H. & Mal. & \fullcirc & \halfcirc & \emptycirc &   \\ 
Bkakria et al.\ \cite{bkakria2020privacy} & Intrusion detection &   & \fullcirc &   & \fullcirc &   & \fullcirc &        & 36ms & 640 KB & S.H. & Mal. & \fullcirc & \halfcirc & \quartercirc & Honesty assumption \\ 
Pine\ \cite{ESORICS:NHPXLWD20} & Intrusion detection &   & \fullcirc &   & \fullcirc &   & \fullcirc & 6k & 665ms & 350 KB & S.H. & Mal. & \fullcirc & \halfcirc & \quartercirc & Mention \\ 
Ren et al.\ \cite{ren2020privacy} & Intrusion detection &   & \fullcirc &   & \fullcirc &   & \fullcirc & 3k & 3.82s & 50 MB & S.H. & Mal. & \fullcirc & \halfcirc & \quartercirc & Honesty assumption \\ 
Han et al.\ \cite{han2020secure} & Intrusion detection &   & \fullcirc &   &   & \fullcirc &   & 24k & $\times$ & $\times$ & TEE & Mal. & \fullcirc & \halfcirc & \halfcirc & Attestation \\ 
Ren et al.\ \cite{ren2019toward} & Intrusion detection &   & \fullcirc &   & \fullcirc &   &   & 3k & 2s & $\times$ & S.H. & Mal. & \fullcirc & \halfcirc & \emptycirc &   \\ 
TVIDS\ \cite{wang2019tvids} & Intrusion detection &   & \fullcirc &   &   & \fullcirc &   & $\times$ & $\times$ & $\times$ & TEE &   & \fullcirc & \halfcirc & \halfcirc & Attestation \\ 
LightBox\ \cite{CCS:DWYZWR19} & Intrusion detection &   & \fullcirc &   &   & \fullcirc &   & 3.7k & 20us &        & TEE & Mal. & \fullcirc & \halfcirc & \emptycirc &   \\ 
Guo et al. (A)\ \cite{guo2018enablingA} & Intrusion detection &   & \fullcirc &   & \fullcirc &   & \fullcirc & 1.6k & 4ms & 2x & S.H. & S.H. & \fullcirc & \halfcirc & \quartercirc & Honesty assumption \\ 
ShieldBox\ \cite{trach2018shieldbox} & Intrusion detection &   & \fullcirc &   &   & \fullcirc &   & $\times$ & 40us &        & TEE & Mal. & \fullcirc & \halfcirc & \halfcirc & Attestation \\ 
SafeBricks\ \cite{poddar2018safebricks} & Intrusion detection &   & \fullcirc &   &   & \fullcirc &   & 18k & $\times$ & 16\% & TEE & Mal. & \fullcirc & \halfcirc & \halfcirc & Third party audit, Attestation \\ 
Snort\ \cite{kuvaiskii2018snort} & Intrusion detection &   & \fullcirc &   &   & \fullcirc &   & 3.4k & $\times$ &        & TEE & Mal. & \fullcirc & \halfcirc & \halfcirc & Third party audit, Attestation \\ 
Alabdulatif et al.\ \cite{alabdulatif2017privacy} & Intrusion detection &   &   & \fullcirc & \fullcirc &   &   &        & 1.3s &        & S.H. &   & \fullcirc & \halfcirc & \emptycirc &   \\ 
BlindIDS\ \cite{canard2017blindids} & Intrusion detection &   & \fullcirc &   & \fullcirc &   & \fullcirc & 3k & 74s & $\times$ & S.H. & Mal. & \fullcirc & \halfcirc & \emptycirc &   \\ 
Trusted Click\ \cite{coughlin2017trusted} & Intrusion detection &   & \fullcirc &   &   & \fullcirc &   & $\times$ & $\times$ & 2x & TEE & Mal. & \fullcirc & \halfcirc & \halfcirc & Attestation \\ 
SGX-Box\ \cite{han2017sgx} & Intrusion detection &   & \fullcirc &   &   & \fullcirc &   & 26k & $\times$ & 11.9\% & TEE &   & \fullcirc & \halfcirc & \halfcirc & Attestation \\ 
Slick\ \cite{trach2017slick} & Intrusion detection &   & \fullcirc &   &   & \fullcirc &   & 10 & 18us & 88\% & TEE &   & \fullcirc & \halfcirc & \halfcirc & Attestation \\ 
Melis et al.\ \cite{melis2016private} & Intrusion detection &   & \fullcirc &   & \fullcirc &   & \fullcirc & 10 & 250ms & 119 B & S.H. & $\times$ & \fullcirc & $\star$\footnote{Client MB separate from cloud MB gets processed info} & \quartercirc & Honesty assumption \\ 
S-NFV\ \cite{shih2016s} & Intrusion detection &   & \fullcirc &   &   & \fullcirc &   & $\times$ & 27us & 8.79x & TEE & Mal. & \fullcirc & \halfcirc & \halfcirc & Attestation \\ 
Shi et al.\ \cite{shi2015privacy} & Intrusion detection &   & \fullcirc &   & \fullcirc &   &   & $\times$ & $\times$ & $\times$ & S.H. & Mal. & \fullcirc & \halfcirc & \quartercirc & Mention \\ 
P2DPI\ \cite{ASIACCS:KCBSPN21} & Data exfiltration, Intrusion detection &   & \fullcirc &   & \fullcirc &   & \fullcirc & 2k & 0.037ms & 840 KB & S.H. & S.H. & \fullcirc & \halfcirc & \emptycirc &   \\ 
PrivDPI\ \cite{CCS:NPLCC19} & Data exfiltration, Intrusion detection &   & \fullcirc &   & \fullcirc &   &   & 3k & 0.15s & 49 B & S.H. & Mal. & \fullcirc & \halfcirc & \quartercirc & Honesty assumption \\ 
CloudDPI\ \cite{li2017clouddpi} & Data exfiltration, Intrusion detection &   & \fullcirc &   & \fullcirc &   &   & 96.6k & 3ms & 28.5 MB & S.H. & S.H. & \fullcirc & \halfcirc & \quartercirc & Honesty assumption \\ 
Lin et al.\ \cite{lin2016privacy} & Data exfiltration, Intrusion detection & \fullcirc & \fullcirc &   & \fullcirc &   &   & 1k & 10us & 1 MB & S.H. & S.H. & \fullcirc & \halfcirc & \emptycirc &   \\ 
Yuan et al.\ \cite{yuan2016privacy} & Data exfiltration, Intrusion detection &   & \fullcirc &   & \fullcirc &   & \fullcirc & 3.2k & 10us & 3x & S.H. & S.H. & \fullcirc & \halfcirc & \quartercirc & Mention \\ 
Embark\ \cite{lan2016embark} & Data exfiltration, Intrusion detection &   & \fullcirc &   & \fullcirc &   & \fullcirc & 100k & 50ms & 4.3x & S.H. & Mal. & \fullcirc & \halfcirc & \emptycirc &   \\ 
PRI\ \cite{schiff2016pri} & Data exfiltration, Intrusion detection &   & \fullcirc &   &   & \fullcirc &   & $\times$ & $\times$ & $\times$ & TEE & Mal. & \fullcirc & \halfcirc & \halfcirc & Attestation \\ 
BlindBox\ \cite{sherry2015blindbox} & Data exfiltration, Intrusion detection, Parental control &   & \fullcirc &   & \fullcirc &   & \fullcirc & 3k & 33us & 2.5x & S.H. & Mal. & \fullcirc & \halfcirc & \quartercirc & Mention \\ 
Ramezanian et al.\ \cite{ramezanian2021parental} & Parental control & \fullcirc & \fullcirc & \fullcirc & \fullcirc &   &   & 100 & 1s & 57.5 KB & S.H. & S.H. & \fullcirc & \fullcirc & \quartercirc & Mention \\ 
\end{tabular}
} 
\caption{\protect Details of all TLS middleboxes found in our literature search.  
Latencies and communication corresponds to processing one packet on all rules.
See caption of Table \ref{tab:individual-nonmiddlebox} for legend.}
\label{tab:individual-middleboxes}
\label{tab:individual-middlebox}
\end{table*}

\subsection{On the inclusion of privacy-preserving deep packet inspection}
\label{sec:tls-middlebox}

Prior analyses of content moderation in end-to-end encryption \cite{rahalkar2022sok,cdt,mayer2019content} have either been content-agnostic or considered content moderation only in the setting of social media or a general E2EE messaging service.  The inclusion of privacy-preserving deep packet inspection in the modern content moderation debate is rare, in part because the focus as of late has been on applications rather than moderation of the general public by Internet Service Providers (ISPs), though internet infrastructure services still perform both voluntary and mandated content removal (e.g.\ \cite{ruddock2021widening,prince2022blocking,bradshaw2018politicization}).

Privacy-preserving deep packet inspection meets all our criteria of E2EE content moderation: it occurs over an E2EE channel (web traffic encrypted with Transport Layer Security, or TLS \cite{TLS}), it seeks to identify specific problematic content (e.g. unauthorized data exfiltration 
or intrusion detection)
and it achieves at least partial client privacy, revealing no additional information about non-problematic content to the moderator (aside from false positives).
It often also provides server privacy for the specific rules, patterns, or data being to be matched against the encrypted traffic.

These methods are also used in privacy-preserving parental control systems, occupy a curious space in between corporate moderation and social media moderation:
The goals of parental control tend to be more aligned with social media goals: blocking content concerning specific topics, e.g. sexual or drug-related.
However the technology used tends to be more highly related to the corporate network monitoring methods---we found many papers for which the technical scheme was self-described for use in both settings \cite{sherry2015blindbox,jia2022encrypted,USENIX:GAZBW22,fan2020group,lai2021practical,canard2017blindids,canard2021towards,li2017clouddpi,lan2016embark,CCS:NPLCC19,sherry2015blindbox}.

Criticisms of deep packet inspection---privacy-preserving or otherwise---also echo the current content moderation debate for E2EE.
A 2017 survey by O'Neill et al.\ \cite{oneill2017tls} investigated the general public's support of TLS proxies (without privacy preservation) and found that more than 70\% were at least somewhat concerned that TLS proxies could be used by hackers or governments.  90\% agreed that browsers should notify users of TLS proxies.  However, there was support for uses in many specific settings: with notification, 80\% believed the use of TLS proxies was permissible by companies for company devices, 69\% said the same for elementary schools, and 64\% for universities.
This makes the problem context extremely important:
Not all content moderation is for social media conversations among adults.

Thus, we unite these two areas of the literature.
The current debate over child safety and misinformation in E2EE has lessons to learn---both helpful and cautionary---from privacy-preserving content moderation of TLS-encrypted network traffic.

\section{Problem contexts: Three examples}
\label{app:case-studies}

This appendix contains three brief examples of how the problem context changes the landscape of feasible detection, response, and transparency mechanisms.

\textbf{Child safety protections.}
Child safety incorporates a host of different topics that broadly seek to protect children from online harms, especially those of a sexual nature.
Levy and Robinson identify seven ``harm archetypes,'' distinguishing between offender-to-offender CSAM sharing, offender-to-victim grooming, live streaming of child abuse, non-CSAM communication between offenders (individual or group), consensual child-to-child indecent image sharing, and viral image sharing (in which CSAM is sent to shock or offend) \cite{levyRobinson}.  Child trafficking is also frequently included in the discussion of child safety, e.g. \cite{borrelli2020non,ncmecissues,uksafetyreport}.
Different interventions apply to different harms: Comparing images in messages against a list of known CSAM held by a clearinghouse such as the National Center for Missing and Exploited Children (NCMEC) will be much more effective for catching offender-to-offender image sharing and viral image sharing \cite{levyRobinson,saferio,googlehash,cybertiplinedata,national2016,iwfhash}, whereas textual and metadata analysis is more suited for detecting offender-to-child grooming or enticement \cite{bours2019detection,lykousas2021large,mladenovic2021cyber}.
The tension between privacy and child safety has been recognized for years, though no resolution has emerged (e.g. \cite{kardefelt2020encryption,ncmecE2E,ukSafetyTechWinners,levyRobinson,biop,eff2021,limniotis2021cryptography}).
The same privacy-safety tradeoff occurs many forms of content moderation under encryption, but child safety is an especially important topic due to the horrific scale and type of the problem: NCMEC received more than 29 million reports of child sexual abuse material, in 2021 alone \cite{ncmec2021}, and the magnitude of online sexual harms to children have grown significantly in the last several years \cite{pozniak2020child}.
For more information on content moderation for child safety, we direct the reader to \cite{googleNcmecThorn,lee2020detecting,guerra2021detecting,levyRobinson}.

In 2021-22, the U.K. ran the ``Safety Tech Challenge'' (UK STC, \cite{ukSafetyTechWinners}) in which companies built various kinds of child safety-focused content moderation systems for E2EE environments; these are described alongside the rest of our literature search.

\textbf{Moderation of hate and harassment.}
Hate and harassment is broadly defined as persistent action toward an individual or group that is meant to cause emotional harm to the target, including causing fear of physical or sexual violence \cite{citron2014hate,thomas2021sok}.  This category itself incorporates a wide variety of behaviors; the recent work of Thomas et al.\ \cite{thomas2021sok} identifies seven categories of hate.  The category most relevant to this work is \emph{toxic content}, e.g. hate speech, sexual harassment, or threats of violence.

For hate specifically, word filters for hate speech have been criticized both for being easy to evade \cite{grondahl2018all}
and for misunderstanding the context when a naively problematic word is being used in a positive way
\cite{MalZam2018,warner2012detecting,fortuna2018survey}.
Many ML classifiers aim to detect hate speech, however they tend to have low accuracies and low agreements (and moreover, humans also have low agreement when it comes to identifying hate speech) \cite{akhtar2020modeling,DBLP:journals/corr/RossRCCKW17}.
As such, automatically detecting hate speech under E2EE would have a high false positive rate.
Moreover, for the kinds of toxic content under discussion, a user is in the loop who does not want the content and could report it.
Automated classification may still have a role to avoid placing the ``burden of responsibility [on] individual, isolated users'' \cite{doi:10.1080/1369118X.2016.1153700} (see also \cite{doi:10.1177/1461444814543163,morozov2013save}), however, in this case it may be more appropriate to perform the classification in a fully client private way.

\textbf{Detecting data exfiltration.}
Many corporations use ``TLS middleboxes'' to perform various services, including detection of intruders and attempts at data exfiltration.
Although the majority of companies today do this in a non-privacy preserving manner
\cite{de2020survey},
there are some middleboxes which perform the detection in a partially client private way (see Section \ref{tab:litsearch}).  These usually use searchable encryption or trusted hardware to ensure that only positive detections are revealed to the middlebox; non-matches remain as private as the underlying cryptography provides.  The exact privacy and detection properties of these middleboxes vary, although partial client privacy is near-universal.  

\section{Attacks on perceptual hash functions}
\label{sec:phf-attacks}

Recent works have demonstrated effective attacks on PHFs.  
In addition to evasion attacks, these works have also demonstrated partial inversion attacks, as well as targeted collision attacks.  We describe each attack in turn:

\paragraph{Evasion}
For the setting of matching via PHFs in content moderation, one of the most serious attacks is evading detection by creating an image that is highly similar to one on the list of problematic content, but has a different hash \cite{jain2021adversarial,DBLP:conf/ccs/HaoLJ021,struppek2021learning,krawetz2021}.
This setting is concerning because the entire motivation for using PHFs rather than CHFs is the reduced false negative rate, thus, effective evasion attacks significantly lower the benefits of using PHFs while retaining the cost of a high FPR.
Jain et al.\ \cite{jain2021adversarial} show both white-box and black-box evasion attacks on a variety of PHFs including PDQ and pHash.
Hao et al.\ \cite{DBLP:conf/ccs/HaoLJ021} show black-box attacks on pHash and a more robust variant known as Blockhash.
Struppek et al.\ \cite{struppek2021learning} demonstrate evasion attacks against NeuralHash along with prototype code.
And Krawetz\ \cite{krawetz2021} shows a proof-of-concept evasion attack against PhotoDNA.
The black-box attacks are especially concerning since they do not require any knowledge about how the algorithm actually works.
This implies that these attacks are viable even in server-private settings.
In settings that require some interactivity in order to compute the function, the server may be able to rate-limit clients who are computing the function too much, but given the frequency with which images and messages are sent in online communication, this form of rate-limiting will also likely interfere too much with normal communication.  We expect black-box attacks to be feasible in the vast majority of content moderation scenarios.

\end{savenotes} 

\paragraph{Finding hash preimages}
Another attack is inversion of hashes.
Two independent proof-of-concept attacks have shown the feasibility of inverting hashes in PhotoDNA \cite{athalye2021,krawetz2021}, yielding a somewhat blurry and distorted version of an image which hashes to a particular known value.
Although these attacks have not been demonstrated yet within the more formal research literature, the initial results demonstrate the importance of server privacy with respect to highly illegal content.

\paragraph{Target hash collisions}
Finally, another line of attacks on PHFs create collisions with target hashes, e.g. \cite{dolhansky2020adversarial} for pHash and other open-source hashes, and \cite{struppek2021learning,ygvar2021} for NeuralHash.
In the setting of Apple's CSAM detector, researchers and others posed concerns that if a hash in a CSAM list becomes known, target collision attacks could plant innocuous images on someone's device that would trigger a CSAM detection \cite{biop,cdt,USENIX:KulMay21,newamerica,claburn2021apple}.  Attackers or protestors could also attempt to overwhelm Apple's human content moderation resources by triggering many adversarially-created matches that match CSAM hashes but are not themselves CSAM.
\label{sec:adversarial-ml}
Note that for matching via lists, adversarially induced false positives presuppose that a client has knowledge of at least one hash on the list.  It is unclear whether this assumption is reasonable in practice.  Adversarial attacks against ML classifiers are in some sense easier; they often require only black-box access to the classifier, rather than knowledge of a confidential list.

We suggest that the privacy loss for an induced collision is not as impactful as a ``true'' false positive -- the sender could choose not to do so and avoid the detection with significantly higher probability.
However, this does not extend to users who might unwittingly receive or forward adversarially-modified messages.  This could, for instance, be used to plant matching material on a target device that would flag the images and potentially open up the target to future investigation.
Beyond privacy, adversarially induced false positives could also be used by malicious actors, activists, or others to overwhelm the detection system or make it useless.
Platforms should be prepared to decide how to detect whether particular users are sending adversarially-induced false positives, determine whether it is feasible to separate these from standard false positives, and potentially modify terms of service to prevent attempts at sending massive amounts of adversarially-created false positives.

\section{Common cryptographic tools for content moderation under E2EE}
\label{sec:crypto}
\label{sec:crypto-methods}
\label{app:crypto-methods}
In this section we briefly describe the cryptographic tools frequently used by the detection mechanisms and provide references for further reading.

\paragraph{Private Set Intersection}

The typical setting for Private Set Intersection (PSI)  \cite{garimella2021oblivious,pinkas2020psi,pinkas2019spot,pinkas2018scalable,pinkas2014faster} is for two parties Alice and Bob to hold secret sets $A$ and $B$ respectively.  PSI allows either Alice, Bob, or both to learn the intersection $A \cap B$ without Bob learning $(A \setminus B)$ or Alice learning $(B \setminus A)$.

PSI is frequently used in exact or perceptual matching to find the intersection of the server's private list $C$ with a client's message $\{m\}$.  The latest PSI schemes are quite fast and PSI schemes specialized for membership testing have low communication complexity.

\paragraph{Searchable Encryption}

Searchable Encryption (SE) \cite{bosch2014survey,curtmola2006searchable,boneh2004public,baek2008public} is an umbrella term combining searchable symmetric encryption (SSE) \cite{curtmola2006searchable} and Public-key Encryption with Keyword Search (PEKS) \cite{boneh2004public,baek2008public}.
The specifics of the scheme vary widely, but SE typically allows Alice to encrypt a list of documents $L$, where each document $D \in L$ is a list of words, in such a way that a designated keyholder Bob can perform a search over a ciphertext to identify or reveal documents $D \in L$ that contain Bob's word $w$.
SE schemes typically have pre-specified leakage in the form of either \emph{index leakage} (leaking information about $L$), \emph{search pattern leakage} (which leaks information about $w$ to Alice), or \emph{access pattern leakage}, leaking information about the relationship between multiple queries $w_1$, $w_2$, and $L$.

The leakage inherent to searchable encryption schemes requires careful evaluation for each scheme to ensure it is compatible with server privacy and partial client privacy; there are known classes of attacks on searchable encryption \cite{kellaris2016generic,cash2015leakage,islam2012access}.

\paragraph{Homomorphic and Functional Encryption}

Both homomorphic encryption and functional encryption consider a ``data owner'' Alice, and an ``evaluator'' Bob.
In both, Alice has the keys necessary to encrypt and decrypt ciphertexts, and Bob has a separate ``evaluation key'' that allows him to manipulate the ciphertext in specific ways, without (necessarily) learning Alice's underlying plaintext.

Homomorphic encryption \cite{acar2018survey,gentry2009fully} allows Bob to perform limited computation on ciphertexts, without learning the result himself:
Partially homomorphic encryption allows the addition of ciphertexts, that is, if $c_1$ and $c_2$ are ciphertexts for $x_1$ and $x_2$ respectively, then there is an addition protocol $\texttt{Add}$ such that $\texttt{Add}(c_1, c_2)$ yields a ciphertext for $(x_1+x_2)$.
Somewhat homomorphic encryption has a similar protocol $\texttt{Mult}$ for multiplication that may be used a limited number of times; fully homomorphic encryption allows unlimited use of $\texttt{Mult}$.
Depending on the scheme, the evaluation key necessary to compute $\texttt{Add}$ and $\texttt{Mult}$ may be a ``key'' as we normally think of them, or it may be the case that ciphertexts can be added and multiplied without having any key at all.

Functional encryption \cite{boneh2011functional} takes this idea a step further: the evaluation key $e$ was generated from a specific secret value $k$ and function $f$.  Bob can use $e$ to compute $f(x, k)$ with access only to $e$ and a ciphertext of $x$.
Among other uses, this allows functional encryption to emulate homomorphic encryption, but also allows specialized decryption (for example revealing to Bob whether $x=k$, and revealing no other information).

Homomorphic encryption (especially fully homomorphic encryption) and functional encryption tend to be slower operations (seconds rather than milliseconds) but are still practical for some settings.

\paragraph{Multi-party computation}

Multi-party computation (MPC) \cite{evans2018pragmatic,boneh2020graduate,katzLindell} generically refers to any method that allows multiple parties $P_1, \ldots, P_N$ to compute a function output $f(x_1, \ldots, x_N)$ without learning anything aside from the output (in particular, without learning anything about the inputs $x_1, \ldots, x_N$.
Sometimes it is used to refer to specific techniques such as secret-sharing \cite{gmw,bgw,spdz}, garbled circuits \cite{yao}, or similar frameworks \cite{aby}; other times it can also refer to generic public-key and symmetric-key protocols run between at least two parties.
In this work we use MPC to refer to any cryptographic protocol that is not one of the other specialized techniques described here.

\paragraph{Trusted Execution Environments}

Trusted execution environments (TEEs) \cite{zheng2021survey,mofrad2018comparison,costan2016intel} like Intel Software Guard eXtensions (SGX) \cite{sgx} provide an isolated encrypted area of memory known as an \emph{enclave}, such that the data within that region cannot be accessed by other software running on that hardware, and SGX can attest that the correct software is running.

Mainly in corporate network monitoring, a common paradigm for the client or gateway forwards the decryption key for the encrypted channel directly to the SGX enclave---out of reach by the service provider itself---and all the desired network functions (e.g. traffic analysis or detection of exfiltrated secrets) take place within the enclave (see e.g. \cite{coughlin2017trusted}).  If the content moderation detection code is running within the enclave and all other information remains unaltered, this exactly meets our definition of partial client privacy (assuming one believes the guarantees of TEEs in general).  TEEs can also enact server-private code, since the client can be denied access to read any encrypted rules.

TEEs have two major downsides: the first is that they require specialized hardware that may not be an option in most content moderation settings.  The second, and more serious, is that TEEs have been heavily criticized for having privacy-crippling side channel attacks via timing, cache, energy, and speculative execution that are capable of recovering encryption keys \cite{nilsson2020survey,chen2019exploitable,van2018foreshadow,chen2018sgxpectre,lindell2018security,gotzfried2017cache}.

\section{Client privacy}
\label{sec:client-privacy}
In this section we elaborate on the privacy issues inherent to different settings of client privacy under different detection paradigms.

\label{sec:privacy-settings}
A significant part of the modern debate on child safety content moderation in E2EE concerns the definition of E2EE and to what extent its guarantees are or are not violated by various detection and response mechanisms.  After analyzing the literature, we see several approaches with conflicting privacy guarantees:

\paragraph{Full client privacy}
The most privacy-preserving approach is to perform the entire pipeline, detection \emph{and response}, on the client's device, with no automated message sent to the server or a moderator.
Any detection mechanism that preserves full client privacy, from matching to machine learning, avoids the problem of leaking false positives to the server.
Many client-side E2EE spam filters meet this requirement
as do many misinformation ``tiplines'' in WhatsApp \cite{martins2021fact,kazemi2021tiplines,meedan2020one,bagade2020kauwa,melo2019whatsapp}, and Apple's nudity classifier for underage accounts in iMessage \cite{appleCommunicationSafety}.
User reporting with message franking removes one part of the deniability guarantee (see Section \ref{sec:message-franking}) but the confidentiality of the message holds unless one of the ends of the message deliberately reveals it.

Full client privacy does not remove ``slippery slope'' questions of whether the scheme could be altered in the future; a small tweak to client-side code would, for most applications, allow the detection to be sent to the server instead.
However, a similar (though more obvious) tweak would allow most E2EE applications to exfiltrate \emph{all} user data to the server; we rely on a variety of technical and non-technical means to detect such a change (see Section \ref{sec:transparency}).

If one is to perform content moderation in E2EE, this is the most privacy-preserving option.

\paragraph{Exact matching (partial client privacy)}
In exact matching, one party---often the server---has a list of problematic content (usually stored as cryptographic hashes).  The server learns whether any of the client's content matches with the list exactly (see Section \ref{sec:exact-matching}), often accomplished by using Private Set Intersection or other multi-party computation.
In principle, the match could also be performed on the client side, however, the literature mostly contains works that achieve full client privacy in that setting.

The exact matching paradigm carves out an important exception to the E2EE confidentiality guarantee: it only holds against non-matches.
However, this method avoids the tricky issue of false positives:
Although false positives are theoretically possible using exact matching, common cryptographic hash functions would only expect to reach a collision with probability $2^{-128} \approx 10^{-38}$,
meaning if 7.5 billion WhatsApp messages are sent per day, even for a list of a billion elements with distinct hashes, it would take longer than the age of the universe to reach a single false positive in expectation.
This category is still vulnerable to the surveillance and slippery slope concerns described in Section \ref{sec:surveil},
but it avoids the privacy issues inherent to schemes with a higher false positive rate.

\paragraph{Predicate/policy exact matching (partial client privacy)}
Some systems, especially seen in corporate monitoring and parental control,
use Searchable Encryption to perform exact matching anywhere in a packet (e.g. it would find a match for ``nana'' in the word ``banana'').
The technologies used to achieve this vary in their leakage.
Schemes based on order-preserving encryption have known attacks revealing message content 
and typically do not meet the standard confidentiality guarantee of E2EE even on non-matches.
Other schemes, based on searchable encryption, have different specified leakage.  These schemes must be examined for privacy leakage on a case-by-case basis.

The privacy issues present in these schemes typically do not involve false positives, but rather involve the cryptosystem itself.

\paragraph{Perceptual matching (partial client privacy)}
In these schemes, one party (typically the server) holds a list of \emph{perceptual} hashes of problematic content.
Similar to exact matching, the server and client perform a protocol to determine whether the perceptual hash of the client's content appears on the server's list of problematic content, and, in partially client private systems, the server learns the result of the match.

This setting begins to significantly degrade the privacy guarantees of E2EE:
the false positive rates of modern perceptual hash functions are in the range from $10^{-3}$ \cite{jain2021adversarial} to $10^{-8}$ \cite{apple2021ThreatModel}.
The cryptographic tools for these systems typically increase the false positive rate only a negligible amount (on par with the amount for exact matching); nearly all of the false positive rate arises from the perceptual hash itself.
Unlike exact matching, this does begin to erode the privacy guarantees of E2EE severely: using the same number of 7.5 billion messages per day, this corresponds to between 4.5 million and 135 false positives per day.

Furthermore, in addition to the false positive problem, the surveillance problems remain.  The false positive problem adds an additional difficulty: if a PHF-based matching system was to be deployed, we believe the approximate false positive rate should be disclosed as a matter of transparency to allow users to make informed choices on the privacy properties of the chat services they use.
We also suggest research into means of verifying the aggregate detection rate in Section \ref{sec:transparency}.
In that section we also suggest methods for cryptographically (and non-cryptographically) addressing appeal and redress.

To our knowledge, no research has been done on the distribution of false positives, but naively we would expect the false positives to be unevenly distributed in the distribution of sent messages.
We call for more research on perceptual hash functions, both to develop more accurate and precise PHFs and also to understand the distribution of false positives.

This setting was precisely the matter at issue in Apple's automated CSAM detector \cite{bhowmick2021apple}.
Weighing the tradeoff between the significant privacy loss of these systems, their surveillance risk, and the horrific acts of child abuse they aim to stop is a policy tradeoff that is informed, but not determined, by this analysis.

\paragraph{ML classification (partial client privacy)}

The accuracy of ML classification varies widely based on the context-specific task and the classifier itself.
Classifiers for content moderation tasks like nudity detection, misinformation, and child enticement achieve accuracies between 70\%-97\%
\cite{hor2021evaluation,mcghee2011learning,garcia2018pornographic,hunt2022monitoring,mladenovic2021cyber,morris2012identifying,pendar2007toward,bogdanova2014exploring}
The common consensus seems to be that at least for now, machine learning approaches have higher false positive rates than perceptual hash functions for the most serious categories of problematic content like CSAM \cite{levyRobinson}.

The privacy impacts on E2EE are extreme, potentially leaking one in every 10-100 benign messages to the server or moderator, potentially leaking hundreds of millions of false positives per day if deployed on the scale of WhatsApp.
This remains true even if the classification is performed client-side on the plaintext of the sender or receiver's device.
For ML classification to be compatible with E2EE, we strongly recommend that either full client privacy be used, or the that significant improvements be made in the classification methods.

\bibliographystyle{ACM-Reference-Format}
\bibliography{bib/abbrev0, bib/abridgedcryptobib, bib/extrabib}

\end{document}